\documentclass{article}
\usepackage[utf8]{inputenc}
\usepackage[linesnumbered,ruled,vlined]{algorithm2e}
\usepackage{graphicx, caption, subcaption}
\usepackage{svg}
\usepackage{amssymb, amsmath}
\usepackage{mathtools}
\usepackage{comment}
\usepackage{ragged2e}
\usepackage{booktabs, multirow, multicol}
\usepackage{color, soulutf8}

\graphicspath{ {images/} }

\title{Introducing norms in the reasoning cycle of emotional aware Jason agents }

\author{\textbf{Daniel Pérez, Estefanía Argente, Elena Del Val, Soledad Valero} \\
Universitat Politècnica de València / Camino de Vera s/n. \\ 46022 Valencia (Spain) \\
Valencian Research Institute for Artificial Intelligence \\
\texttt{\{dapregar, eargente, edelval, svalero\}@vrain.upv.es}
}

\date{}

\begin{document}

\maketitle

\begin{abstract}
Human relationships are complex processes that often involve following certain rules that regulate interactions and/or expected outcomes. These rules may be imposed by an authority or established by society. In multi-agent systems, normative systems have extensively addressed aspects such as norm synthesis, norm conflict detection, as well as norm emergence. However, if human behaviour is to be adequately simulated, not only normative aspects but also emotional aspects have to be taken into account. In this paper, we propose a Jason agent architecture that incorporates norms and emotions in its reasoning process to determine which plan (actions) to execute. The proposal is evaluated through a scenario based on a social network, which allows us to analyse the benefits of using emotional normative agents to achieve simulations closer to real human world.

\textbf{Keywords}: Multi-agent systems, social agents, norms, emotions, reasoning process, behaviour
\end{abstract}

\section{Introduction}
Human societies are defined by a complex amalgam of interactions between individuals \cite{powers2021cooperation}. These interactions can be modelled taking into account the knowledge that an individual has about his environment: laws, social norms, previous interactions, emotions, etc. \cite{kennedy2012modelling} 
Traditionally, when modelling intelligent agents to simulate human behaviour, works have approached it from a normative \cite{marir2019norjade, shams2017practical, viana2015jsan, meneguzzi2015bdi, lee2014n, boissier2013multi, criado2013manea, panagiotidi2012reasoning, alechina2012programming, oren2011acting, dos2011developing} or emotional \cite{taverner2018modeling, alfonso2017toward, Becker10} perspective, keeping these two approaches as separate subjects.

Norms are undoubtedly useful when defining logically explainable situations. They can be used to establish both the trigger condition and the expected outcome  
\cite{vazquez2005organizing}. Therefore, norms are useful to express computationally how an individual may react towards a certain event when working within a controlled situation. Additionally, there are different types of norms that could be used depending on whom they affect, or whom they are promulgated by. For instance, institutional norms \cite{Criado11, Boella08b} are those promulgated by an authority (such as a government) and are expected to be followed by the society no matter the circumstances. In order to do so, obligations, prohibitions and, when required, permissions are established. These norms are usually predefined by code or can be transmitted by a regulatory agent. When an agent decides to comply with a certain institutional norm it might be rewarded in order to reinforce that attitude. Similarly, it can be sanctioned when violating the norm. Social norms \cite{Savarimuthu} arise from society, and can be seen as conventions tacitly acquired between the parts. They are informal, shared rules of behaviour that prescribe what individuals should or should not do, and are followed because of social expectations and possible social sanctions \cite{szekely2021evidence}. Finally, private norms or moral norms represent agent's own values.

But human behaviour is not only constrained by norms. We are not merely rational beings driven by rules and, obviously, we do not expect only rational driven behaviours when interacting with someone else \cite{cohen1981can}. 
Interactions between human beings generate emotions and, consequentially, are oftentimes also triggered by emotions \cite{aureli2002relationship}. 
Much work in the field of psychology has shown that emotions, feelings, motivations, moods and other affective processes are not only linked to our well-being, but also shape our behaviour and drive key cognitive mechanisms such as attention, learning, memory and decision-making \cite{andrade2009enduring, updegraff2004makes, so2015psychology}.
DeWall et al. \cite{dewall2016often} discuss about how the emotional state of a person defines the way it behaves and responds towards an event. For instance, when a person suffers fear from something or someone, its instinct will force them to flee from the source of its fears. The authors call this the emotion-as-direct-causation theory. Additionally, in Baumeister et al. \cite{baumeister2007emotion} also consider the emotion-as-feedback theory, which proposes that anticipated emotions (i.e. which emotions might arise from ones actions) are a usual factor use to choose which behaviour might be the best in order to achieve the desired emotional outcome. For instance, when someone is sad, they usually might perform actions that will try to improve their emotional state. This second theory also implies that people learn to anticipate what actions will result in which emotions. In conclusion, the emotion-as-direct-causation perspective asserts that current emotions guide behaviour and judgement, whereas the emotion-as-feedback perspective asserts that anticipated emotions guide behaviour and judgement which means that, although the default assumption is that emotions are the proximal cause of behaviour and judgement, the anticipated emotions also impact reliably in social behaviour and judgement. Considering affective processes in cognitive and behavioural models can increase the explanatory and predictive power of such models \cite{hasking2017cognitive}.


A lot of work has been done in the field of Multi-Agent Systems (MAS) both through norm enforcing \cite{boella2006introduction, marir2019norjade, shams2017practical, viana2015jsan, meneguzzi2015bdi}, and through the use of emotions \cite{yannakakis2014emotion, taverner2018modeling, alfonso2017toward, Becker10} as independent/non-directly connected variables. Nonetheless, when designing a system that is trying to simulate human behaviours or interact with them, both norms and emotions shall be used together \cite{bourgais2019ben, kollmann2016towards, schaat2015modelling, schaat2017psychologically} in order to achieve more accurate results.

Emotional normative agents will have a relevant application in so-called hybrid social realities \cite{perko2020hybrid}, where humans will be assisted by physical and digital agents, with indistinguishable behaviours. For instance, when developing an intelligent agent intended to interact with humans, such as a recommendation system supposed to persuade humans in order to take the best decision, or a personal assistant, it is vital to make the system as human-like as possible. Other possible usage can be found when designing software intended to predict how an individual or a group of people will react towards an event, such as a natural catastrophe or a fire in a building. In these cases, norms can be used to determine the most logical and/or expected behaviours, whereas emotions can be used to modulate how this logical behaviour should be carried on. For example, the intensity of a response might be modulated depending on the current emotional state of the agent; an agent might decide to violate a norm because her current mood promotes this; or a new social norm might emerge given the emotional reactions towards an unexpected event.

In a previous paper \cite{argente2020normative}, we defined four relationships between norms and emotions that should be considered when designing normative-emotional agents: (1) emotions (more precisely, agent mood and anticipated emotions) must be taken into account in the process of normative reasoning, since current agent mood and the perspective of the society mood (i.e. anticipated emotions) can affect compliance behaviour; (2) compliance with or violation of a norm generates emotions in the agent who performs the action regimented by the norm (e.g. pride when compliance, guilt or shame when non-compliance), but also in the observers of this action (e.g. admiration or reproach, respectively); (3) emotions enforce social norms, since anticipated emotions can help reflecting both empirical expectations (people's beliefs about what others will do) and normative expectations (people's beliefs about what others think that one ought to do)  \cite{szekely2021evidence}; and (4) emotions allow the norms to be internalised, so that certain behaviours accepted by society and/or which positively influence the affective state of the agent, may end up being considered as private norms, seen as principles or concerns of the agent himself. 
Considering these four relationships, an abstract BDI architecture for a Normative Emotional Agent (NEA) was defined. Based on this architecture, here we define a Normative-Emotional Agent architectural proposal for Jason.

The article is structured as follows: in section \ref{sec:MostRelevantProposals}, an in-depth review of the most important works on Normative-Emotional agents in Jason is given. Section \ref{sec:ArchitecturalProposal} describes the Normative-Emotional Agente architectural proposal for Jason. Section \ref{sec:DemoScenario} provides an example of how agents perform their reasoning processes using their normative knowledge and their emotional state. Finally, section \ref{sec:Conclusions} presents conclusions and expected future work derived from this article.

\section{Norms and Emotions in Jason} \label{sec:MostRelevantProposals}
Most of the approaches on normative agents \cite{meneguzzi2009norm, dos2011developing, lee2014n, meneguzzi2015bdi, viana2015jsan} are based on AgentSpeak programming language \cite{rao1996agentspeak}, proposing an extension of Jason with normative issues. Jason \cite{bordini2005bdi} is a well-known agent platform, well documented and easily downloadable from its website.

At the time of writing, there are three versions of AgentSpeak. The first version, AgentSpeak(L)  \cite{rao1996agentspeak}, proposed in 1996, is the first proposal to formalize BDI agents, allowing the abstraction of the intrinsic complication of these systems. Besides being the oldest version, it is also the most widely used and serves as the basis for the other two proposals. AgentSpeak(PL) \cite{silva2011agentspeak}, launched in 2011, allows to represent the beliefs of an agent as probabilistic knowledge of its environment, through Bayesian Networks. AgentSpeak(ER) \cite{ricci2018agentspeak}, proposed in 2018 and included in Jason in September 2021, adds plan encapsulation, i.e. the possibility to define plans that fully encapsulate the strategy to achieve the corresponding goals, integrating both pro-active and reactive behavior.

\begin{figure}
    \centering
    \includegraphics[width=1\textwidth]{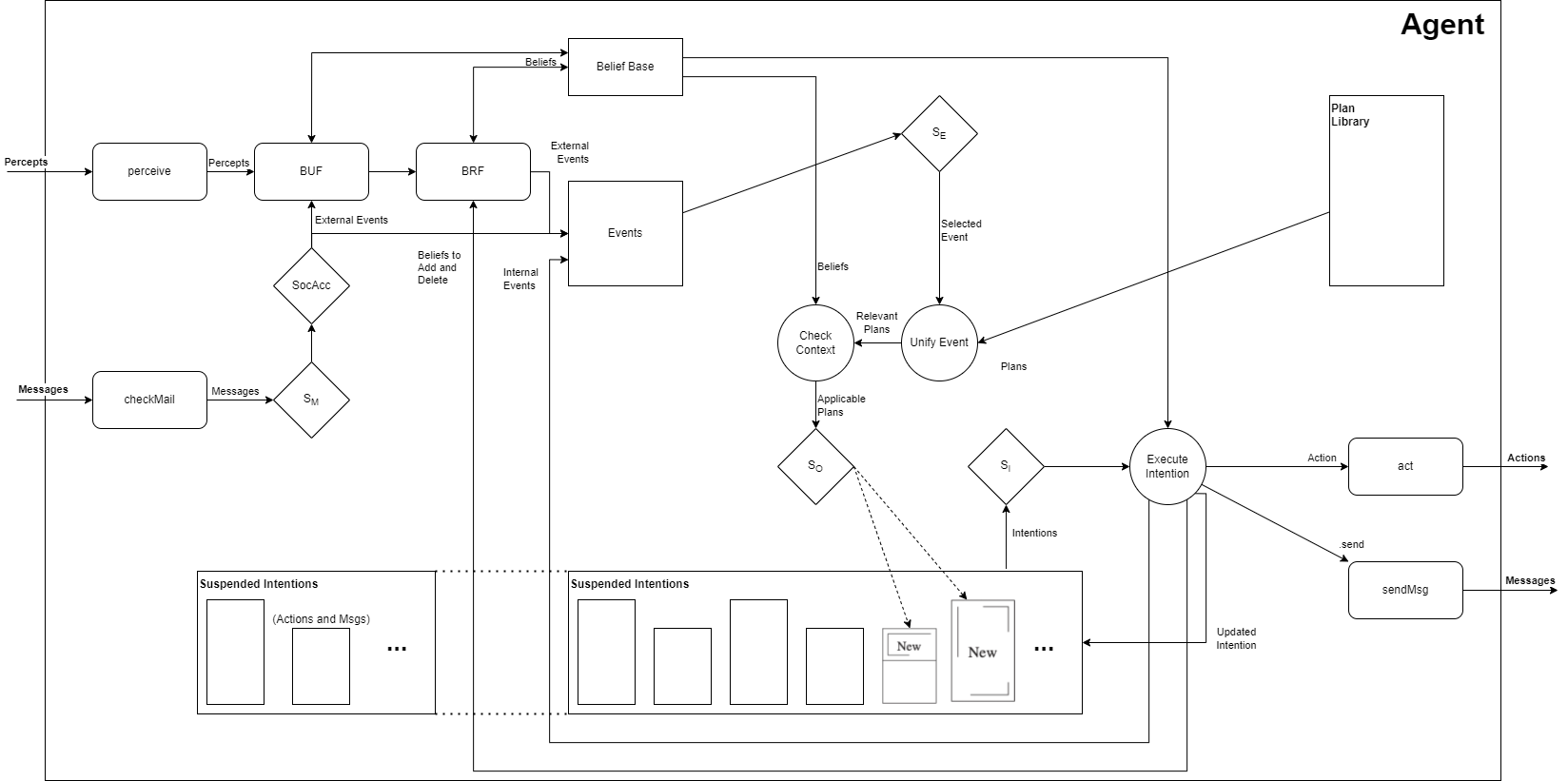}
    \caption{Jason Reasoning Cycle \cite{bordini2005bdi}}
    \label{fig:jason_reasoning_cycle}
\end{figure}

Jason agents are based on believes, goals and plans. The reasoning cycle of an agent in Jason (see Figure \ref{fig:jason_reasoning_cycle}) follows a series of steps \cite{bordini2005bdi}.  First, the agent perceives the environment and uses these perceptions to update its beliefs. Each change in the belief base generates an external event. The agent can also receive messages from other agents which, once accepted, they also generate external events. Using the updated beliefs, the agent can then compute which desires are actually possible to accomplish. This is done by first selecting an event from the current event list. The plans relevant to the event are then chosen from the plan library and, taking into account the current beliefs, the agent determines the plans that are applicable and selects one of them. The selected plan will update the set of intentions. Finally, using some valuation function, an intention is selected and its related plan is executed. Then the intentions are updated, removing the executed plan from the set of intentions and the cycle repeats again.

We will now review the different proposals in Jason / AgentSpeak for both normative agents and emotional agents, and describe how these works have addressed the concept of norm and/or emotion, as well as what relevant modifications they have made to the reasoning cycle of a Jason agent so that these aspects can be taken into consideration.

\subsection{Norms in AgentSpeak}
In this section we analyze the most relevant normative agent papers for AgentSpeak. 

\vspace{0.5cm}

\textbf{IOVIS} \cite{meneguzzi2009norm}
Iovis is an implementation of an AgentSpeak interpreter that includes a meta-level reasoning action library, and plans to process and comply to norms. With Iovis, a developer can deploy agents that change their plan libraries in reaction to norms from the environment, or communicated by other agents, by creating new plans to comply with obligations and suppressing the execution of existing plans that violate prohibitions. Their proposal is implemented in the AgentSpeak(L) language. 

Norms are represented by beliefs with the following structure: 
\textit{norm(Activation, Expiration, Norm)}, where \textit{Activation} and \textit{Expiration} represent the states that will trigger the activation and deactivation of the norm, respectively, and \textit{Norm} $\in$ \textit{obligation(X), prohibition(X)} represents the norm itself, where X represents an action or a state, \textit{obligation(X)} indicates the need for adding a new plan for the specified action, or adding a new goal to achieve the specified state; and \textit{prohibition(X)} prevents the adoption of any plan that executes the indicated action or leads to the indicated state. For simplicity, they assume that norms can activate and expire only once through their life-time. This implies that if the expiration condition of a norm is already true when a norm is adopted, it is assumed to have expired and the norm is ignored. 

When an agent is told a norm, he will incorporate it directly into his set of beliefs. When the activation condition is already true, this means that the norm must be enacted immediately and any contrary behavior must be stopped. For an obligation, this implies that the agent must either react to reach the state of the world (by adding a goal to reach that state) or execute the action specified in the obligation (by adding a new plan with the triggering event of the activation, and the obligated action in its body). For a prohibition, if the activation condition is already true, this means that the agent must refrain from reaching the offending world state or executing the offending action, suppressing plans that may violate the prohibition and discarding any instances of these plans adopted as intentions. Finally, when both the activation condition and the expiration condition are false, the agent must create new plans to enforce compliance as soon as the activation condition is met by adding them to the plan library. 

In this work, authors have focused on the expected actions an agent must undertake in order to comply with norms. Agents created under this architecture are capable of incorporating new norms at runtime. However, aspects such as the decision on whether or not to accept a norm and the verification of its consistency (i.e. it does not conflict with its current norms) have been left for future work. Likewise, once the norm is incorporated into the belief-case, as the necessary plans for its achievement are added, the norm is rigorously followed, with no reasoning mechanisms available to evaluate the possible advantages of skipping it. Additionally, agents lose their normative knowledge once the norm has expired, since all normative plans are eliminated.

\vspace{0.5cm}

\textbf{NBDI} \cite{dos2011developing}
In this paper authors propose a modification in the BDI architecture to obtain normative agents that allow to represent obligations and prohibitions. For this purpose, they propose the following formal specification: \textit{norm (Addressee, Activation, Expiration, Rewards, Punishments, DeonticConcept, State)} where: \textit{Addressee} is the agent or role responsible for norm compliance, \textit{Activation} is the activation condition of the norm, \textit{Expiration} is the deactivation condition of the norm, \textit{Rewards} is the reward associated with norm compliance, \textit{Punishments} is the punishment associated with non-compliance with the norm, \textit{DeonticConcept} indicates whether the norm is an obligation or a prohibition, and \textit{State} represents the set of states regulated by the norm.

In its reasoning cycle, the agent first determines whether the new percept is a norm, in which case it compares it to the norms already stored in its belief base and then updates the belief base accordingly. Next, based on its current beliefs, the agent updates its set of active norms. If the expiration condition has been met, the agent deactivates the norm and stores it as an adopted one. An update of desires and priorities is performed in the next step, without modifying the process described in the reasoning cycle of a BDI agent. After having updated its fact base, desires and priorities, the agent reviews the set of active norms and chooses those that best help it to meet its objectives either by following the norm or by breaking it. In addition, he identifies and resolves any conflicts between selected norms. Next, the agent selects those plans that will allow it to fulfill its desires, taking into account the norms it has selected. Finally, the agent executes the specified actions based on the agent's intention.
 
Agents created under the architecture proposed in this paper are able to incorporate new norms at runtime, as well as to break norms if this would be more favorable for them. Unfortunately, the pseudocode examples found in the paper make use of some functions that, despite being new contributions and having their purpose explained, are not detailed as to how they could be implemented despite being important for understanding how the agent's reasoning process works. 

\vspace{0.5cm}
\textbf{N-Jason} \cite{lee2014n} This work makes use of plan annotations, incorporating through them the different concepts related to the norms, such as the deadline of an obligation, the lifetime of a prohibition, their priority, etc.

When a norm is communicated to it, the agent will first try to recognize whether that norm has the format \textit{$p(t_e, edp, R_p)$} where \textit{p} defines whether it is an obligation or a prohibition, \textit{$t_e$} defines the event that will be generated by the new norm, \textit{edp} defines the deadline and priority of the norm and \textit{$R_p$} defines a set of relevant plans, which may or may not be within the agent's existing plans. If the norm has the expected format, the agent proceeds to modify its event base by adding triggering events as targets achievable by the agent.

If the plans contained in the norm are in the agent's plan library, in the case of an obligation, the event that ensures its fulfillment is added, while in the case of a prohibition, the event is added to the prohibition base for further evaluation. Subsequently, when the agent considers which of its intentions it wishes to fulfill, it will obtain all the possible intentions that it can satisfy and organize them according to their priorities and deadlines, leaving the prohibitions as the intentions with the lowest priority, and additionally using the established deadline to try to execute the obligations that, with lower priority, may be closer to it.

N-Jason agents are able to incorporate new rules at runtime, but they are not able to reason whether the breach of a rule may be advantageous for them, since the normative reasoning only considers the priority and deadline of the triggering events associated with the rules, without taking into account the consequences (i.e., penalties, rewards) of the actions performed.

\vspace{0.5cm}
\textbf{v-BDI} \cite{meneguzzi2015bdi} In this paper the authors propose a normative agent that allows representing mandatory obligations and prohibitions. For this purpose, the following abstract representation of the norm is realized: $\langle D_{a}$\textit{, action, activation, deactivation, id}$\rangle$ where $D_{a}$ indicates the deontic logic of the norm (i.e. obligation or prohibition) and the agent $a$ affected by the norm, \textit{action} indicates the action affected by the norm, \textit{activation} and \textit{deactivation} are beliefs that respectively indicate the activation and deactivation condition of the norm, and \textit{id} reflects the identifier with which the norm is labeled and with which it may be referenced by other norms.

\begin{figure}[ht]
    \centering
    \includegraphics[width=1\textwidth]{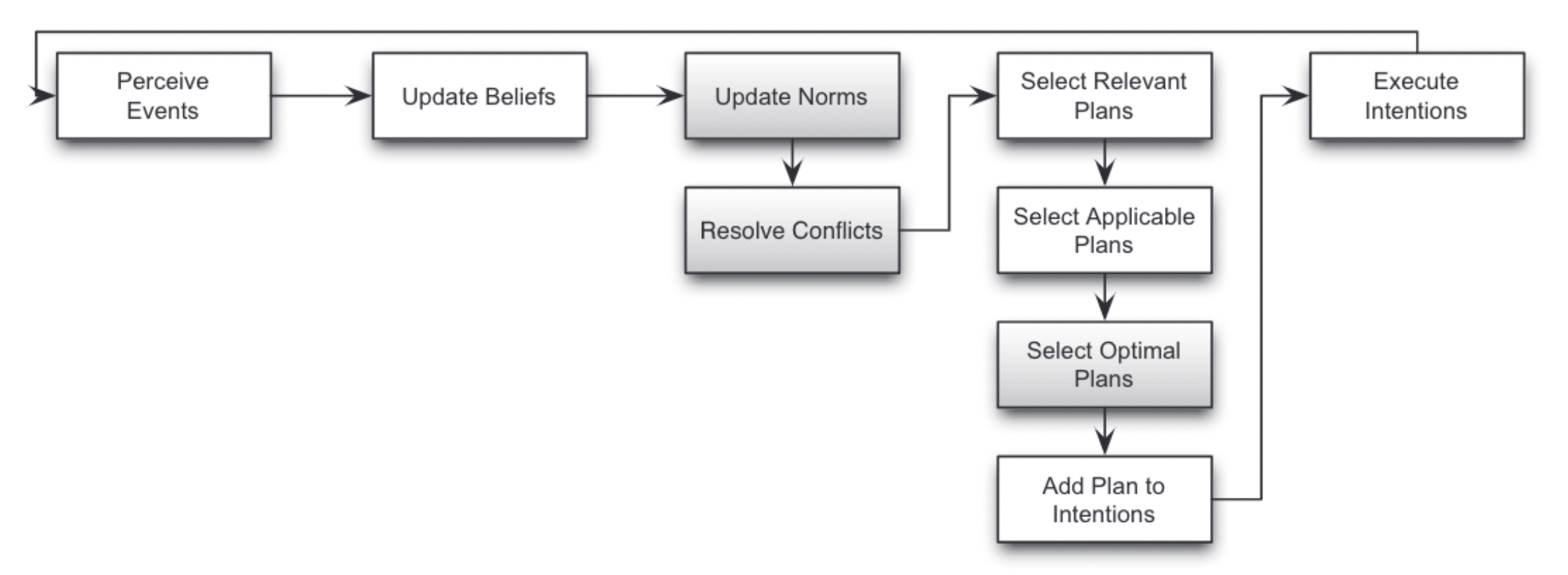}
    \caption{v-BDI reasoning cycle \cite{meneguzzi2015bdi}}
    \label{fig:vbdi_reasoning}
\end{figure}

Figure \ref{fig:vbdi_reasoning} shows the agent's reasoning cycle. Once the belief update has been performed, the agent proceeds to store the new normative beliefs that have been communicated to it. Then, with the intention of resolving conflicts between prohibitions and obligations that may affect the same plan, in case an obligation or prohibition already exists for a plan/state, the agent will ignore any other norm that affects that plan/state when its activation and deactivation conditions overlap. Finally, when the time comes to select the plan to execute, the agent chooses it on the basis of a utility function that is calculated as: the sum of the plan's base utility, i.e., the utility associated with the successful achievement of goals (as instantiated by the plans used to achieve them) and the plan's benefit utility, i.e., the utility associated with the plan's fulfilment of norms, minus the plan's cost utility, i.e., the utility associated with the plan's violation of norms.

Agents created under this proposed architecture are capable of reasoning whether compliance or non-compliance with a norm is favorable to them, as well as of incorporating new norms at runtime. However, due to the way in which norms are adopted, once the expiration period is reached, the normative beliefs are erased, so the agent loses the acquired normative knowledge. Finally, the authors mention the importance of the resolution of conflicts between norms, but the paper itself points out that this aspect has been worked on superficially, due to its complexity, ignoring for example the conflicts that may arise from an ambiguous definition of a norm, something that is likely to happen given that the definition of a norm itself does not have a formally defined structure.

\vspace{0.5cm}
\textbf{JSAN} \cite{viana2015jsan} In this paper the authors propose a normative agent that allows representing mandatory obligations and prohibitions. To represent the norms they make use of the following formal specification: $\langle$\textit{Addressee, Activation, Expiration, Rewards, Punishments, DeonticConcept, State}$\rangle$ (same as the one found in the work of Neto et al. ) where \textit{Addressee} represents the agent responsible for rule compliance, \textit{Activation} and \textit{Expiration} represent respectively the rule activation and deactivation condition, \textit{Rewards} and \textit{Punishments} represent the rewards associated with compliance with the rule and the penalties in case of non-compliance, \textit{DeonticConcept} indicates whether it is an obligation, prohibition or permission and \textit{State} represents the state/event/action affected by the rule.

In this work, it is proposed that agents have the following capabilities: 

\begin{itemize}
    \item Norm awareness: agents identify which norms are active in the environment, and thus, these norms are then assigned to specific agents.
    \item Norm adoption: agents acknowledge their responsibilities towards other agents through internalisation of norms in which their responsibilities are specified.
    \item Norm deliberation: for the purpose of executing a specific norm, an agent needs access to different information, such as the objectives to be hampered by the fulfillment of normative objectives and the objectives that might profit from associated rewards.
    \item Norm impact: following the agents' execution of the norm, agents' objectives are updated. Thereafter, the cycle continues and agents begin to identify other norms that ought be addressed.
\end{itemize}

Although the paper states that agents are capable of performing all the functions described, it does not detail how the reasoning cycle of a Jason agent is modified to incorporate these features, nor does it provide any algorithm that allows us to know how each task is accomplished. It is therefore impossible to know how the rules are managed and how the agent's knowledge is actually modified.

A comparison between the different normative agent proposals in AgentSpeak that we have analyzed is shown in table \ref{tab:normative-agents-comparison}. 

\begin{table}[ht]
\resizebox{\textwidth}{!}{
    \begin{tabular}{|l|lllllll|l|l|}
    \hline
    \multirow{2}{*}{} & \multicolumn{7}{c|}{Normative Concepts Specified} & \multirow{2}{*}{\begin{tabular}[c]{@{}l@{}}Add norms\\ during runtime\end{tabular}} & \multirow{2}{*}{Ignore norms} \\ \cline{2-8}
     & \multicolumn{1}{l|}{Addressee} & \multicolumn{1}{l|}{\begin{tabular}[c]{@{}l@{}}Activation\\ Condition\end{tabular}} & \multicolumn{1}{l|}{\begin{tabular}[c]{@{}l@{}}Expiration\\ Condition\end{tabular}} & \multicolumn{1}{l|}{Rewards} & \multicolumn{1}{l|}{Punishments} & \multicolumn{1}{l|}{\begin{tabular}[c]{@{}l@{}}Deontic\\ Concept\end{tabular}} & \begin{tabular}[c]{@{}l@{}}Affected\\ Actions\end{tabular} &  &  \\ \hline
    IOVIS & \multicolumn{1}{c|}{} & \multicolumn{1}{l|}{\checkmark} & \multicolumn{1}{l|}{\checkmark} & \multicolumn{1}{l|}{} & \multicolumn{1}{l|}{} & \multicolumn{1}{l|}{\checkmark} & \checkmark & \checkmark & \checkmark \\ \hline
    NBDI & \multicolumn{1}{l|}{\checkmark} & \multicolumn{1}{l|}{\checkmark} & \multicolumn{1}{l|}{\checkmark} & \multicolumn{1}{l|}{\checkmark} & \multicolumn{1}{l|}{\checkmark} & \multicolumn{1}{l|}{\checkmark} & \checkmark & \checkmark &  \\ \hline
    N-Jason & \multicolumn{1}{l|}{} & \multicolumn{1}{l|}{} & \multicolumn{1}{l|}{\checkmark} & \multicolumn{1}{l|}{} & \multicolumn{1}{l|}{} & \multicolumn{1}{l|}{\checkmark} & \checkmark & \checkmark &  \\ \hline
    v-BDI & \multicolumn{1}{l|}{\checkmark} & \multicolumn{1}{l|}{\checkmark} & \multicolumn{1}{l|}{\checkmark} & \multicolumn{1}{l|}{} & \multicolumn{1}{l|}{} & \multicolumn{1}{l|}{\checkmark} & \checkmark & \checkmark & \checkmark \\ \hline
    JSAN & \multicolumn{1}{l|}{\checkmark} & \multicolumn{1}{l|}{\checkmark} & \multicolumn{1}{l|}{\checkmark} & \multicolumn{1}{l|}{\checkmark} & \multicolumn{1}{l|}{\checkmark} & \multicolumn{1}{l|}{\checkmark} & \checkmark & \checkmark (?) & \checkmark (?) \\ \hline
    \end{tabular}
}
\caption{Comparison of papers on normative agents in AgentSpeak. The question mark indicates that, although the authors indicate in the paper that the indicated functionality is covered, they do not specify how it has been implemented}
\label{tab:normative-agents-comparison}
\end{table}

\subsection{Emotions in AgentSpeak}
In this section we analyse the most relevant works of emotional agents in AgentSpeak.

\begin{figure}[ht]
    \centering
    \includegraphics[width=1\textwidth]{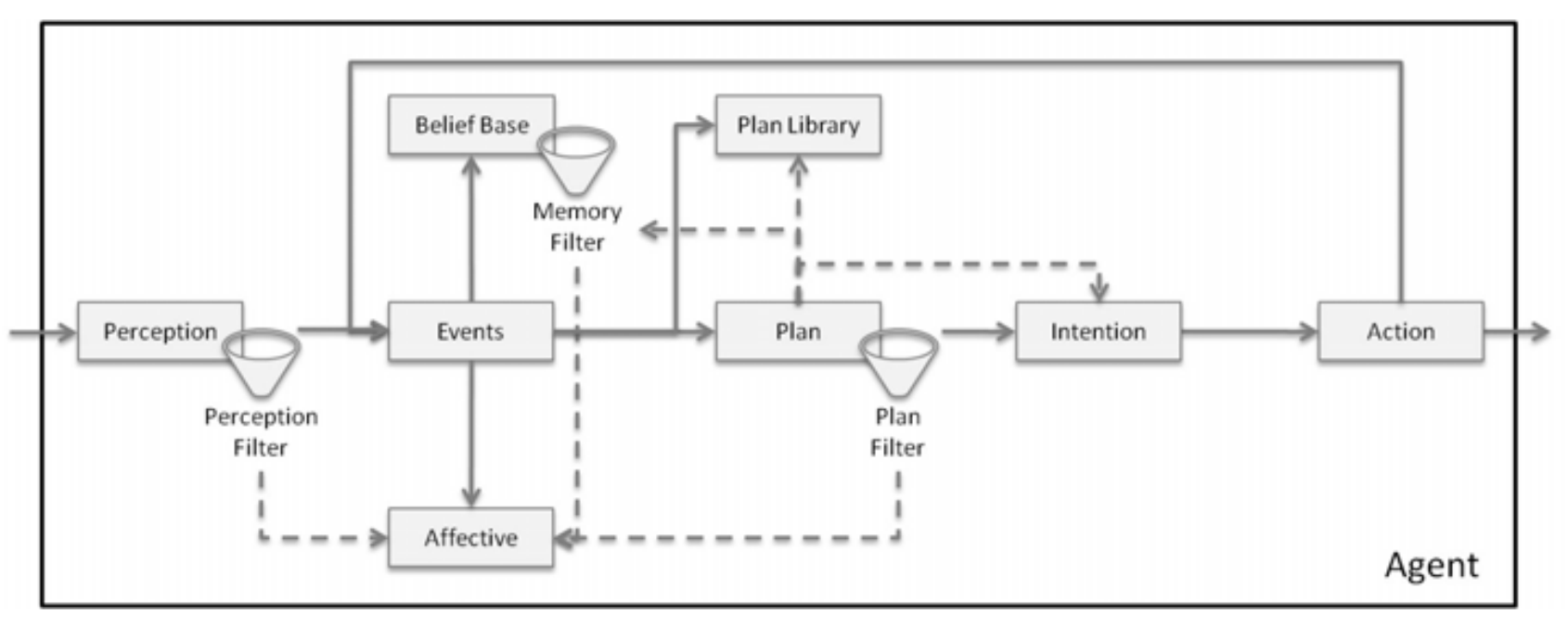}
    \caption{Neto et al. \cite{neto2010construction} reasoning cycle}
    \label{fig:neto2010}
\end{figure}

\vspace{0.5cm}
\textbf{Neto et al.} \cite{neto2010construction} In this paper, the authors propose an extension to AgentSpeak's BDI model through an affective module that allows modifying the agent's cognitive process, affecting its perception, memory and decision-making capabilities (see Figure \ref{fig:neto2010}). This affective module consists of three subcomponents that allow it to represent the personality, the emotional state and the different emotions affecting the agent, respectively.   

To define personality, the authors make use of the Big Five model, also called the OCEAN model \cite{mccrae1992introduction} (Openness, Conscientiousness, Extroversion, Agreeableness and Neuroticism), which measures these five personality traits, assigning a value in the range [-1,1] to each factor. 

To define the emotional state, the authors make use of the PAD model \cite{mehrabian1996pleasure}, named after its three factors (Pleasure, Arousal and Dominance). This model proposes a three-dimensional vector space in the range [-1,1], with each factor associated with an axis, representing by a point in this space the emotional state based on the influence of each factor.

Finally, emotions, which represent short-lived states or immediate reactions to events, are implemented by the authors through the OCC \cite{ortony1990cognitive} model, which represents 22 different emotions, each evaluated in the range [0,1].
,  

Thanks to this affective module, the authors modify the reasoning cycle of a Jason agent by introducing three filters. The first one (perception filter) is used to modify the perception of the agent, including the emotions associated with the perception and the intensity of those emotions. With this filter, if the intensity of the emotion is below a preset threshold, or if the emotion is not aligned with the emotional state (i.e. they do not have the same sign, positive or negative), then the perception is discarded. 
 

The second filter (memory filter) extracts information from the agent's belief base (used to represent his memory of past events) and selects only those events that are congruent with the agent's current affective state, and can also filter on the basis of the emotional intensity of the stored events.

Finally, the third filter (plan filter) assigns a positive or negative mark to the plans, depending on whether or not they are aligned with the agent's current emotional state. Those plans with a positive mark are candidates for execution, while those with a negative mark are discarded.

\vspace{0.5cm}
\textbf{Gluz et al.} \cite{gluz2014probabilistic} In this paper authors propose the use of BDI agents that, through probabilistic computation, are able to determine their emotional state according to the plans they execute, also taking into account their environment. To achieve this, they make use of the OCC model to represent emotions, and AgentSpeak(PL).

To do so, the developer must design as many plans as necessary for each type of emotion to be considered. Plans must be designed to calculate the utility of the execution of each action, and additionally, plans must be designed to calculate the intensity of the emotion that would be elicited if the action were executed. The result is agents capable of performing emotion appraisal.

\begin{figure}[ht]
    \centering
    \includegraphics[width=1\textwidth]{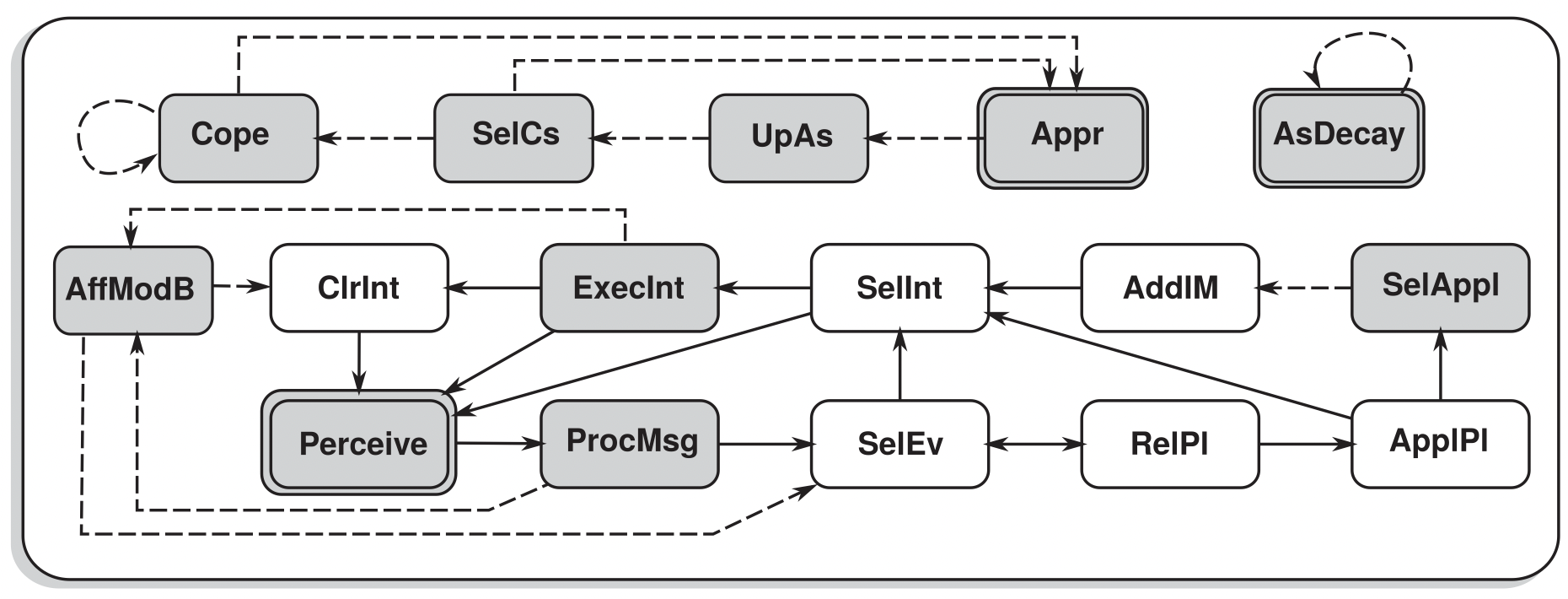}
    \caption{GenIA$^3$ reasoning cycle \cite{alfonso2017toward}}
    \label{fig:genia}
\end{figure}

\vspace{0.5cm}
\textbf{GenIA$^3$} \cite{alfonso2017toward} \cite{taverner2018modeling}. The authors of this paper propose an extension of the reasoning cycle of an AgentSpeak BDI agent (see Figure \ref{fig:genia}). This extension consists of two new cycles, called affective cycle and temporal dynamics cycle, thanks to which it is possible to represent the affective state of an agent.

The affective cycle begins with the \textit{Appr} step, where the \textbf{appraisal} process is performed on the basis of the \textit{concerns, personality and probabilities} represented through the agent's \textit{beliefs}. This step further determines whether an event is affectively relevant to the agent and generates the necessary \textit{appraisal variables} to reflect this. Then, in the step \textit{UpAs}, the affective state is updated based on these \textit{appraisal variables}. After that, the step \textit{SelCs} checks whether new behaviors need to be created in the agent after this change, and also verifies which \textit{coping strategies} are applicable. The \textit{Cope} step performs the necessary tasks to execute the corresponding \textit{coping strategies}. The intentions generated through these \textit{coping strategies} are added to the end of the agent's intention base, which contains the intentions of both the rational cycle and the affective cycle.

The cycle of temporal dynamics (\textit{AsDecay}) allows to regulate the affective state of the agent. When the agent begins its execution, it does so in an affective state of equilibrium, i.e., it does not feel any sufficiently significant emotion. With the passage of time its affective state varies according to the emotions that may be generated, so that the objective of the function \textit{AsDecay} is to regulate this change in order to try that the agent gradually recovers the affective state of equilibrium.

The rational cycle, meanwhile, begins with the \textit{Perceive} step, where the agent observes changes in the environment to update its \textit{beliefs}. The agent then proceeds to update its belief base to reflect these changes, unifying the agent's perception with its beliefs, creating or deleting \textit{beliefs} as appropriate. After updating its belief base, the agent executes the \textit{belief revision function} step. This process consists of several steps in Jason's original setup, but the authors simplify it in the schema by calling this step \textit{ProcMsg}. After that, the agent proceeds to select the event it wants to resolve in the \textit{SelEv} step. It then searches the plan library for all relevant plans (\textit{RelPl}), filters those that are applicable to the event (\textit{AppPl}) and selects one of them (\textit{SelAppl}). After that, the agent performs a similar process to select which intention it is going to solve through the steps \textit{AddIM, SelInt, ExecInt, ClrInt} in which it will add the necessary intentions based on the new beliefs (\textit{AddIM}), select the one it wants to solve (\textit{SelInt}), execute the intention (\textit{ExecInt}) and remove it from the list of intentions (\textit{ClrInt}). Finally, in the step \textit{AffModB}, it proceeds to synchronize the state between the rational cycle and the emotional cycle. This synchronization consists in the creation or elimination of beliefs that have been modified in the rational cycle and that could be necessary in the affective cycle.

\begin{figure}[ht]
    \centering
    \includegraphics[width=0.75\textwidth]{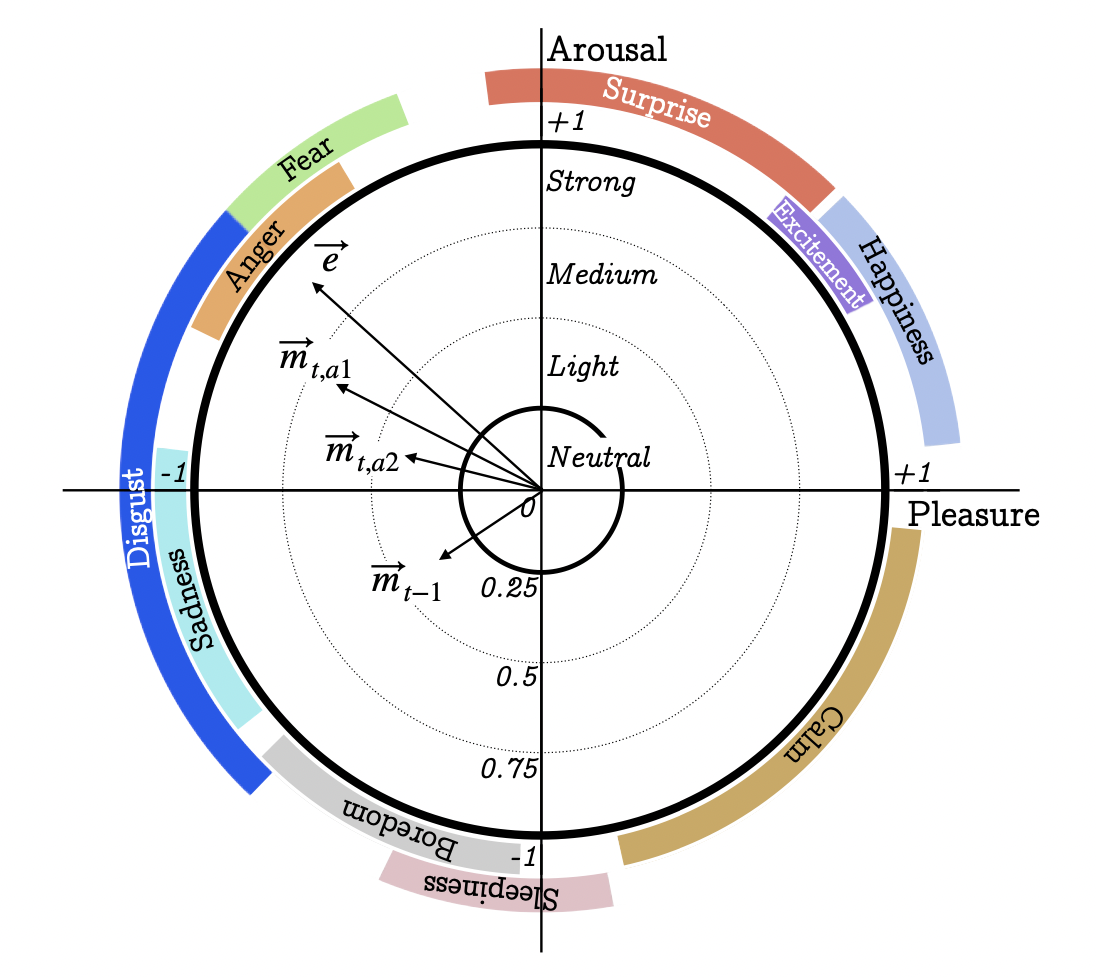}
    \caption{GenIA$^3$ mood representation using the pleasure-arousal dimension  \cite{taverner2021fuzzy}}
    \label{fig:genia_emotions}
\end{figure}

To represent emotions, the authors use the Pleasure and Arousal dimensions of the PAD model. For this purpose, they propose a two-dimensional vector space (see Figure \ref{fig:genia_emotions}) in which each dimension can take a value in the range [-1.0,1.0], representing by means of a vector the affective state and its intensity. Furthermore, making use of the agent's affective memory, the way in which the agent's affective state varies depends both on the current affective state and on the new emotions that he/she perceives.

This proposal, as in Neto et al.'s work, makes use of the OCEAN, OCC and PAD models to represent the agents' personality, emotions and emotional state, respectively. Thus, agents are able to modify their behavior by taking into account how the events they perceive influence their emotional state as long as they are affectively relevant. Additionally, the agent is able to modify its behavior, selecting the actions to execute according to its affective state and its level of rationality, the latter being a parameter of the agent's personality that indicates how much its emotions will influence its decision making.

As a comparison between the proposals analyzed, the table \ref{tab:emotional-agents-comparison} shows which theoretical models are used to represent personality, emotions and affective states, and also the types of emotions that have been considered, according to whether they are immediate emotions (those generated as a reaction towards an event) or anticipated emotions (those that might be generated based on pasts experiences), according to the classification made by Loewenstein-Lerner \cite{4274}.

\begin{table}[ht]
\resizebox{\textwidth}{!}{
    \begin{tabular}{|l|lll|l|l|}
    \hline
    \multicolumn{1}{|c|}{\multirow{2}{*}{}} & \multicolumn{3}{c|}{Theoretical Models Employed} & \multicolumn{1}{c|}{\multirow{2}{*}{\begin{tabular}[c]{@{}c@{}}Considers\\ Immediate\\ Emotions\end{tabular}}} & \multicolumn{1}{c|}{\multirow{2}{*}{\begin{tabular}[c]{@{}c@{}}Considers\\ Anticipated\\ Emotions\end{tabular}}} \\ \cline{2-4}
    \multicolumn{1}{|c|}{} & \multicolumn{1}{c|}{Personality} & \multicolumn{1}{c|}{Emotions} & \multicolumn{1}{c|}{\begin{tabular}[c]{@{}c@{}}Affective\\ State\end{tabular}} & \multicolumn{1}{c|}{} & \multicolumn{1}{c|}{} \\ \hline
    Neto et al. & \multicolumn{1}{l|}{OCEAN/BIG FIVE} & \multicolumn{1}{l|}{OCC} & PAD & \checkmark &  \\ \hline
    Gluz et al. & \multicolumn{1}{l|}{} & \multicolumn{1}{l|}{OCC} &  & \checkmark &  \\ \hline
    GenIA$^3$ & \multicolumn{1}{l|}{\begin{tabular}[c]{@{}l@{}}Adaptive\\ By default OCEAN/BIG FIVE\end{tabular}} & \multicolumn{1}{l|}{\begin{tabular}[c]{@{}l@{}}Adaptive\\ By default OCC\end{tabular}} & \begin{tabular}[c]{@{}l@{}}Adaptive\\ By default PAD\end{tabular} & \checkmark &  \\ \hline
    \end{tabular}
}
\caption{Theoretical models employed and types of emotions considered in each proposal for the Loewenstein-Lerner \cite{4274} classification for the state-of-the-art emotional agents in Jason}
\label{tab:emotional-agents-comparison}
\end{table}

\section{Relationship between norms and emotions}
The relationship between emotions and norms are critical in computer environments in which agents must emulate human behaviors or social groups (i.e., social system simulations, simulated human teams, or virtual agents engaging with or interacting with humans). In such environments, it is critical that agents display emotional reactions regarding norms being followed or violated, to increase realism. In our previous work \cite{argente2020normative}, we established four relationships between norms and emotions. In this section we explain how our proposal tackles each relationship.

\begin{figure}[ht]
    \centering
    \includegraphics[width=0.5\textwidth]{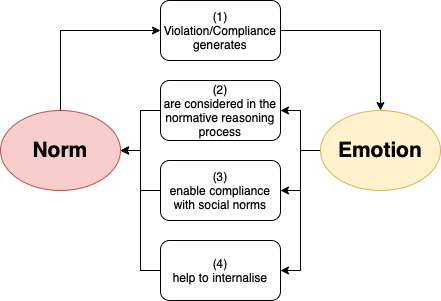}
    \caption{Relations between norms and emotions}
    \label{fig:norm_emotions_relationship}
\end{figure}

\begin{enumerate}
    \item \textbf{Norm violation/ norm compliance generates emotions.} By complying with a norm or violating it, an agent could generate emotions (positive or negative) in its affective state. Von Scheve et al. \cite{VonScheve2006, Fix06} define that compliance with norms generates positive emotions in both the agent and the observers of such compliance, while non-compliance with the norm would generate negative emotions. 
   
    Thus, for instance, when an agent performs an action regulated by a norm, such as the prohibition to \textit{Beat} another agent, if he complies with the norm, he will elicit a positive emotion in himself and in the observers. On the other hand, if it violates the norm, it will generate a negative emotion in itself and in the observers. In the case of wanting to model an aggressive agent, the developer could define in the agent that the action of hitting will generate a positive emotion in him. In this case, the agent will generate an emotion that is the opposite of the one that would be generated in society.
    Therefore, when an agent executes an intention, if this intention is associated with an active norm, the agent will have to update his emotional state, creating a new belief with the elicited emotion for its subsequent synchronisation in the affective cycle. Similarly, when an agent perceives another agent's compliance or non-compliance with a norm, it must update its affective state, and inform the other agent about the emotion that has elicited its action.
    In our proposal, we will allow the designer to define the emotions that will be elicited in the case of compliance or non-compliance with a given norm, respectively. We will also allow the designer to indicate which roles are observers of a norm and, therefore, are emotionally affected by other agents' compliance or non-compliance with the norm. This is achieved thanks to a component called \texttt{list\_of\_affected\_roles} which is part of the norm definition.
    When selecting the intention to carry out, the agent should make use of its current affective state as well as the emotions that can be elicited on the basis of compliance or non-compliance with a norm, in order to determine whether or not it wants to break a norm. In order to establish in what way emotions are considered, we have employed the proposal of Fochmann et al. \cite{fochmann2019happy}. This work concludes that the value of the affective state is inversely proportional to the interest in complying with a norm. That is, an affective state with a high numerical representation increases the probability of noncompliance, and vice versa. For example, if the agent is simulating the action of driving, when approaching a red traffic light he has the possibility of stopping using the action \textit{Brake}, which has the associated norm of obligation when the proposed condition occurs. In case the driver whose behavior is to be replicated is euphoric, the probability that the agent would decide to follow the rule (and therefore choose the \textit{Brake} action) would decrease to the point of breaking the rule by not performing the action.
   
   \item \textbf{Emotions enable compliance with social norms.} When the agent is informed of the emotions elicited in society as a result of performing one of its actions, it will proceed to update its affective state (creating affective beliefs) based on the information received. Thus, when the agent is going to select the intention he wishes to fulfil, he will try to choose those that please society as long as his intention is to maximise the reward. For example, when the agent enters a room he can perform the action \textit{Greet} and the action \textit{Sit}. This case reflects a common social situation, where even though there is no legislation requiring us to greet other people when we arrive in a room, we do so because society has tacitly decided that it is appropriate to do so. Replicating this behaviour, when the agent performs the action \textit{Greet} he observes that he receives a positive affective response, while when he chooses the action \textit{Sit} without having previously greeted, he receives a negative affective response. After N executions the agent will recognise that, by performing the action greeting before sitting down, he will receive more positive affective responses, thus increasing the probability that he will perform this behaviour and, therefore, comply with the social norm.
    
    \item \textbf{Emotions help to internalise the norms.} Similar to the previous point, when the agent receives affective responses from society to its actions, the agent updates its affective state to reflect how they affect it. In case the affective responses to an action are always the same (a pattern is produced) the agent includes this information in its belief base through a \textit{normative belief}. For example, if the agent were to perform the action \textit{Greet}, which might not be regulated by a norm or might not be known, the agent would observe a positive affective response in society. In case of having generated an affective response the agent would create in the same step a new low-relevance \textit{normative belief} indicating the need to comply with this action. If this situation were to be repeated N times in the future, the agent, based on his experience, would reinforce the relevance of the created normative belief, thus guaranteeing compliance with the social norm.
\end{enumerate}

\section{Normative-Emotional Agent architectural proposal for Jason}\label{sec:ArchitecturalProposal}



A normative-emotional agent (NEA) is one that, making simultaneous use of norms and emotions, modifies its behavior to adapt to its environment \cite{argente2020normative}. In this section, we propose an extension of the reasoning cycle of the Affective Jason agent proposed in the GenIA$^3$ architecture \cite{alfonso2017toward} to include norms and thus be able to model the relationships between norms and emotions \cite{argente2020normative}. First, a NEA agent will be formally described using AgentSpeak. Next, the agent's reasoning process will be depicted, detailing its reasoning cycles. Then, the functions and transition rules updated in this proposal will be described. And finally, the extension of the Jason language required by our proposal will be described.  

\subsection{Definition of a Normative Emotional Agent (NEA)} \label{sec:OperationalSemantics}

Since AgentSpeak is the foundation on which Jason's operational semantics is built, it is necessary to extend it to include the affective and normative components needed to represent a NEA. The configuration of an agent in AgentSpeak is defined by the tuple $\langle$\textit{e ag, C, M, T, s}$\rangle$. In \cite{alfonso2017toward}, an extension to include affective components in an agent using AgentSpeak was proposed by defining the tuple $\langle$\textit{ag, C, M, T, Mem, Ta, s, ast}$\rangle$. Based on the latter tuple, we propose here a redefinition of its components, as shown in figure \ref{fig:nea_configuration}, and detailed below:

\begin{figure}[ht]
    \centering
    \includegraphics[width=\textwidth]{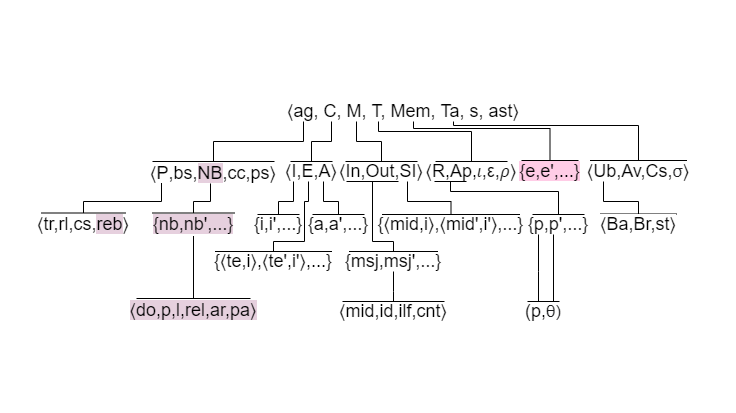}
    \caption{NEA configuration. New components highlighted according to GenIA$^3$ configuration}
    \label{fig:nea_configuration}
\end{figure}

\begin{itemize}
    \item $ag$ represents the agent. It is defined by a set of beliefs (\textit{bs}), a set of plans (\textit{ps}), a set of concerns (\textit{cc}), a personality (\textit{P}) and a set of normative beliefs (\textit{NB}). Both $NB$ and $P$ have been extended in our proposal.  
    \begin{itemize}
        \item Personality \textit{P} includes: the personality traits \textit{tr} (represented in our proposal with the OCEAN model \cite{mccrae1992introduction}); the ``level of rationality" \textit{rl} (used to define to what extent the affective state will affect the agent's decision making and represented in our model with an integer value in the range [0.0, -1.0]. The lower the value, the greater the relevance of the affective state in the agent's decision making); the set of coping strategies \textit{cs} (used to relate a particular affective state to a set of actions with which to generate intentions for the agent, represented in our model by beliefs); and a trait of defiance \textit{reb} (used to define the degree to which the agent will decide to follow the rules, represented as a numerical value in the range [0.0,1.0]. The higher the value, the greater the probability that the agent will decide to break the rules).
        \item The set of normative beliefs \textit{NB} is used to represent the normative knowledge of the agent. These beliefs are represented as:
        \subitem - a deontic operator °textit{do} where \textit{do} $in$ \{\textit{obligation, prohibition}\}, which indicates whether it is a rule of obligation or prohibition.         
        \subitem - a normative plan \textit{p} such that \textit{p} $\in$ \{\textit{ps}\}, i.e., \textit{p} is a plan that represents an action previously known to the agent.
        \subitem - a limit \textit{l} indicating the agent's maximum cycle up to which he must respect the prohibition, if applicable, or comply with the corresponding obligation. 
        \subitem - a relevance \textit{rel}, which makes it possible to define the relevance of the rule, according to the number of times it has been reinforced.
        \subitem - a set of roles \textit{rel} that indicates which roles will be emotionally affected by compliance or non-compliance with the rule.
        \subitem - a pre-appraisal \textit{pa} containing an estimate of how compliance with the norm will affect the agent emotionally. Emotions are represented (in all the components of the agent that make use of them) using the Pleasure and Arousal dimensions of the PAD model. Therefore, to store the pre-appraisal value, a vector in the format [X,Y] is used, where X and Y are two values in the range [-1.0,1.0].
      \end{itemize}
    \item $C$ represents the agent's current circumstances using the tuple $\langle I, E, A\rangle$ where: $I$ is a set of intentions $\langle i, i', ...\rangle$ (each intention $i$ is a stack of partially instantiated plans); $E$ is a set of events $e$ where $te$ is a triggering event and $i$ is an intention (a stack of partially instantiated plans in the case of internal events, or an empty intention $\top$ in the case of events). ...\$ in which $te$ is a triggering event and $i$ is an intent (a stack of plans in the case of internal events, or an empty intent $\top$ in the case of external events); $A$ is a set of actions $\{a,a',...\}$ to be performed by the agent in the environment.
    \item $M$ represents the agent's communication parameters by using the tuple $\langle In, Out, SI\rangle$ where: $In$ and $Out$ correspond respectively to the agent's message inboxes and outboxes, implemented as the set $\{msj,msj',...\}$. Each message ($msj$) is a tuple $\langle mid,id,ilf,cnt,cnt\rangle$ in which $mid$ contains the message identifier code, $id$ the identifier of the agent sending the message, $ilf$ the illocutive force, and $cnt$ the message content; $SI$ is the set of intentions suspended due to message processing, represented with the set of tuples $\{langle mid,i'\rangle, \langle mid',i'\rangle, ...\}$ in which $mid$ contains the identifier of the message that caused the intention to be suspended and $i$, which contains the suspended intention.
    \item $T$ contains the agent's temporal information in the current reasoning cycle, represented with the tuple $\langle R, Ap, \iota, \varepsilon, \rho \rangle$ in which: $R$ represents a set of relevant plans $\{p,p',...\}$ (for the event being solved), $Ap$ represents a set of applicable plans (i. e. relevant plans for which their context is believed to be $true$ and finally $\iota$, $\varepsilon$ and $\rho$ representing an intention, an event and an applicable plan (respectively) to be considered during the execution of a single reasoning cycle.
    \item \textit{Mem}, added in the GenIA$^3$ configuration, contains a set of events $\{e, e', ...\}$ that have been affectively relevant to the agent. This set of events is stored as a memory and is updated in the appraisal process. To this set of events are added those that are significant for the agent at the normative/affective level, represented as the affective responses that the agent receives from society (which are added in the processing of incoming messages) and as the agent's own affective reactions of pride or shame generated by the agent's own compliance or non-compliance with a norm (which are added after executing a norm-affected plan).
    \item \textit{Ta}, defined in the GenIA$^3$ configuration through the tuple $\langle Ub, Av, Cs, \sigma\rangle$, represents the information used by affective processes. 
    \begin{itemize}
        \item \textit{Ub} is the tuple $\langle Ba, Br, st \rangle$, which contains the agent's affective temporal information. Specifically, when at some step in the agent's reasoning cycle it is required to add or remove beliefs to the agent's belief base, instead of adding them directly, they are added to the agent's affective temporal information, so that they can be processed later in the \textit{AffModB} step. Thus, \textit{Ba} represents the beliefs to be added, \textit{Br} represents the beliefs to be removed, while \textit{st} contains the label of the reasoning cycle step in which it has been established that these beliefs were required to be added or removed. In our proposal, normative beliefs will also be taken into consideration herein. 
        \item \textit{Av} contains a set of numerical values for the appraisal variables: desirability, likelihood, expectedness, controllability and causal attribution in the current affective cycle. These variables have a value in the range [0,1] except expectedness, which has a value in the range [-1,1].
        \item \textit{Cs} contains the set of coping strategies that will be executed in the current affective cycle. The way coping strategies are represented is by making a relation between a particular affective state (represented through beliefs) and a set of actions that will generate the intentions to be included in the agent's active intentions.
        \item $\sigma$ represents the agent's current affective state. It contains a set of variables {v, v', ...} where each variable contains a numerical value representing the intensity or the presence or not (in the case of a bivalent variable) of either an emotion category (e.g., sad, happy, angry), an appraisal variable (e.g., desirability, controllability), or a mood dimension (e.g., the dimensions of the PAD model \cite{mehrabian1996pleasure}). In this paper we use the PAD model to represent affective state.
    \end{itemize}
    \item \textit{s} is a label indicating the current step in which the agent is in the transitions diagram (see Figure \ref{fig:transition_states}) of its normative reasoning cycle, where \textit{s} $\in$ \textit{\{Perceive, ProcMsg, SelEv, RelPl, ApplPl, SelAppl, AddIM, SelInt, ExecInt, ClrInt, AffModB\}}. The normative reasoning cycle is explained in detail in section \ref{sec:normative_reasoning_cycle}.
    \item \textit{ast} is a label indicating the current step of the affective reasoning cycle the agent is in, where ast $\in$ \{Appr, UpAs, SelCs, Cope\}. The affective reasoning cycle is explained in section \ref{sec:affective_reasoning_cycle}.
\end{itemize}

Making use of the semantics used in the work of Alfonso et al. (being similar to that used in \cite{vieira2007formal}), we will make use of subscripts to represent the components of the agent. For example, to represent the agent's rebelliousness feature it would be denoted as $ag_{P_{reb}}$.

\subsection{NEA Reasoning Cycles}
\label{subsec:normative_plans_creation}
\begin{figure}[ht]
    \centering
    \includegraphics[width=1\textwidth]{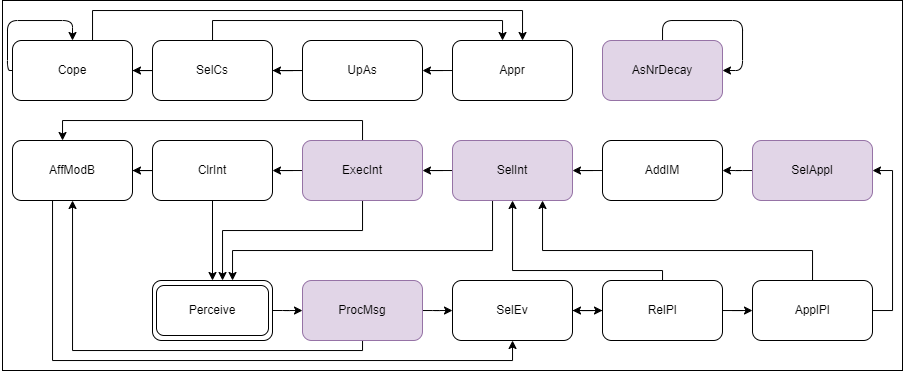}
    \caption{Transition states within one reasoning cycle. Colored steps and dashed arrows represent changes over Alfonso et al. \cite{alfonso2017toward}}
    \label{fig:transition_states}
\end{figure}

As discussed above, for our proposal we have extended the reasoning cycle of an affective agent in Jason from Alfonso et al. \cite{alfonso2017toward} to include the necessary normative processes with which to implement an emotional normative agent. Figure \ref{fig:transition_states} shows the steps that comprise the agent's reasoning cycles, showing in more detail the functions that encompass these cycles and that allow us to transition between the different steps of the agent's cycles, as well as the relationship between the different cycles. 

The normative reasoning cycle comprises the following steps: perceiving from environment (Perceive), processing received messages (ProcMsg), selecting an event from the set of events (SelEv), retrieving all relevant plans (RelPl), checking which of those are applicable (ApplPl), selecting one particular applicable plan, i.e. the intended means (SelAppl), adding the new intended means to the set of intentions (AddIM), selecting an intention (SelInt), executing the selected intention (ExecInt), clearing the executed intention (ClrInt) and synchronising the agents beliefs between the normative cycle and the affective cycle (AffModB). 

The affective reasoning cycle comprises the following steps: performing the appraisal process for perceived events that are affectively relevant (Appr), updating the affective state based on the appraisal performed (UpAs), selecting the most appropriate coping strategy based on the changes in the affective state (SelCs) and executing the selected coping strategy, inserting the corresponding intentions (Cope).

Finally, the cycle of temporal dynamics comprises: updating the affective state to reduce its intensity and reducing the relevance of norms (AsNrDecay).

\begin{figure}[ht]
    \centering
    \includegraphics[width=1\textwidth]{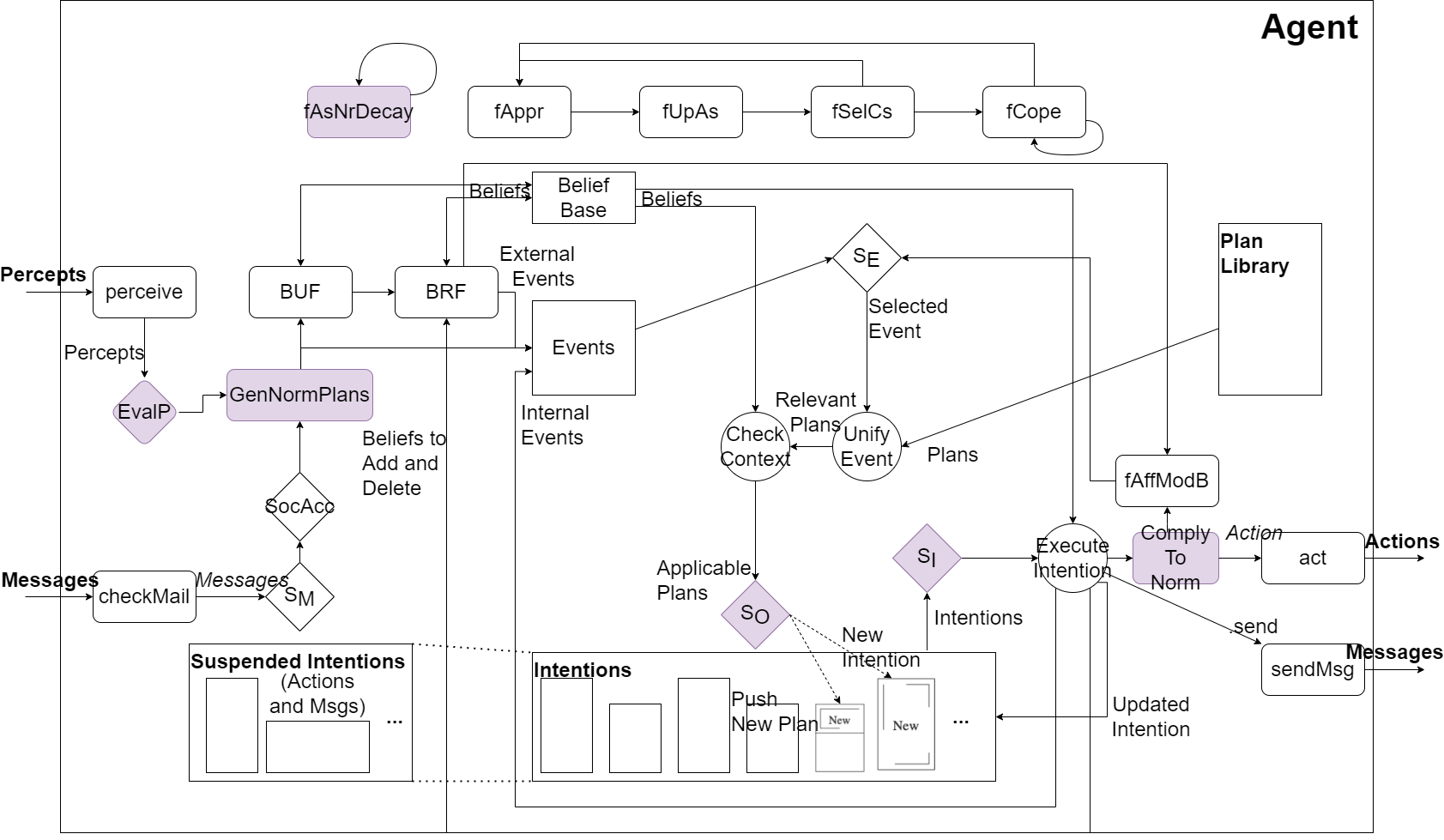}
    \caption{Modified Jason Reasoning Cycle. Colored steps represent changes over Alfonso et al. \cite{alfonso2017toward}. Rectangles represent the main architectural components that determine the agent state. Rounded boxes and diamonds represent functions that can be customised by programmers, being the latter used to represent selection functions (i.e. they receive a list of items and return one). Circles represent essential parts of the interpreter that can not be modified.}
    \label{fig:jason_reasoning_cycle_modified}
\end{figure}

Figure \ref{fig:jason_reasoning_cycle_modified} shows the steps of the reasoning cycles in more detail, including the functions that are used, both within each step and to transition from one step to another. In section \ref{sec:transitionRules} we will detail the functions and transition rules that we have modified and/or extended in our proposal. 

We will now discuss the three NEA reasoning cycles.

\subsubsection{Normative Reasoning Cycle} \label{sec:normative_reasoning_cycle}
\textbf{Perceive Step} \\
The normative rational cycle begins at step \textit{Perceive} with the agent's relationship to its environment, either through the function \textit{perceive} or through the function \textit{checkMail} (see Figure \ref{fig:jason_reasoning_cycle_modified}). Where the agent finds new perceptions by observing its environment, the function \textit{perceive} will collect them and send them to the selection function \textit{EvalP}. When the agent receives the information directly, through messages from other agents, the function \textit{checkMail} will collect these messages and send them to the selection function \textit{select message (\textit{$S_M$})}.

\vspace{0.5cm}

\textbf{ProcMsg Step} \\
At this stage of the reasoning cycle the agent advances to the step \textit{ProcMsg}, which as we have already seen can be reached from two different functions. If it has been reached through the function \textit{perceive}, the agent will execute the function \textit{EvalP}. This function has been modified so that in the event of receiving new norms they can be processed by the agent (i.e. those perceptions observed as new possible norms that do not exist in the belief base) generating a new set of perceptions that it will send to the function \textit{GenNormPlans}. In case the step has been reached through the \textit{checkMail} function, the agent will select one of the messages using the selection function \textit{select message (\textit{$S_M$})}. It will then determine whether the message is socially acceptable (using the selection function \textit{SocAcc)} and, if so, it will pass the content to the function \textit{GenNormPlans}. The function \textit{GenNormPlans} will check in both cases whether the agent's new beliefs are normative in nature and, if they are, the agent will proceed to create the necessary normative plans, contained in the beliefs themselves. After that, it will update the rest of the beliefs with the function \textit{belief update function (BUF)} and will execute the function \textit{belief revision function (BRF)}, which in the default configuration is not modified and we have not needed to extend it.

\vspace{0.5cm}

\textbf{SelEv Step} \\
Once the message has been processed, the agent advances to the step \textit{SelEv}, where it will select the event it wishes to resolve in the current reasoning cycle. As there may be several events pending to be resolved simultaneously, the agent selects which one it will deal with during this reasoning cycle with the function \textit{select event ($S_E$)}. In our proposal we have not needed to modify this selection function.

\vspace{0.5cm}

\textbf{RelPl and ApplPl Steps} \\
Next, the agent advances to the step \textit{RelPl}, in which it must choose a list of relevant plans (i.e. the agent, taking into account its know-how, chooses all those plans that may be relevant, checking which ones unify with the selected event) to subsequently filter this list with the applicable plans (i.e. those in which their context is evaluated to \textit{true} taking into account the agent's beliefs in that reasoning cycle). In our proposal we have made no modifications to this process, since the components responsible for it are essential parts of the interpreter as shown in Figure \ref{fig:jason_reasoning_cycle_modified}.

\vspace{0.5cm}

\textbf{SelAppl Step} \\
Having selected a set of possible applicable plans, the agent must select one of them for execution. This will imply that, once selected, the agent will have the \textit{intent} to follow the course of action determined by the plan, so it will eventually be 'stored' in the agent's set of \textbf{Intentions}. The selected plan is called as 'intended means' since its course of action will be the 'means' that the agent will use to fulfill its 'intention'. This choice is made by the selection function \textit{select option} ($S_O$). In our proposal we have modified this function, which by default is implemented to select plans in the order in which they have been communicated to the agent (or written in its source code), so plans that have normative beliefs associated with them have higher priority. Dignum et al. \cite{dignum2000towards} define that ``The preference ordering of norms is based on a preference of social benefit of a situation, while the preference ordering of obligation is based on the punishment when violating the obligation". Based on this, our agent defines that obligations and prohibitions have a higher priority than the rest of the plans, given that for the agent the punishment associated with the non-compliance of a norm is more important. The criteria used for the ordering is as follows:

\begin{enumerate}
    \item Plans reflecting obligations have the highest priority and are ordered in descending order according to how close they are to meeting their maximum execution cycle.
    \item Plans reflecting prohibitions have the second highest priority and are sorted in descending order according to how close they are to meeting their maximum execution cycle. Since agents can comply with prohibitions as long as they do not execute actions affected by prohibitions, they may end up complying with the rule either voluntarily (by deciding to follow the prohibition's normative plan) or involuntarily by executing other actions not affected by prohibitions. This is why a higher priority is given to obligations, since the only way to ensure compliance is for the agent to voluntarily decide to execute the action obliged by the norm.
    \item The rest of the possible plans are inserted according to the criteria proposed in the original configuration of the agent (order of communication to the agent). In this way, if any action has an associated prohibition, the agent's intention to execute the plan that corresponds to the prohibition is greater and it would not execute the prohibited action.
\end{enumerate}

In our proposal we generate two plans for the agent to contemplate compliance or non-compliance with each rule, so that once ordered the agent evaluates with an utility function that takes into account (amongst other parameters) his affective state and his rebelliousness trait whether he will choose the plan that implies compliance with the rule or the one that implies non-compliance.

\vspace{0.5cm}

\textbf{AddIM and SelInt Steps} \\
Having selected an intended mean, the agent adds it to its set of \textit{intentions} in the AddIM step. As with events and plans, the agent may have at the same time several intentions competing to be the focus of the agent's attention, so it should select only one to attend to during the current reasoning cycle. This is achieved with the intent selection function \textit{SelInt}. In Jason's default configuration, the selection function is based on 'round-robin' planning, which implies that all intentions receive equal attention and are selected in order of insertion. Since rules may need to be fulfilled before a certain reasoning cycle, they are inserted at the top, in order of selection order of the plan that generates the intent, so that they are processed before the rest of the agent's intents.

\vspace{0.5cm}

\textbf{ExecInt and ClrInt Steps}
Once an intention has been selected, the agent proceeds to execute it in the ExecInt step. Since this step is an essential part of the interpreter it cannot be extended, so it has been necessary to create a new function called \textit{ComplyToNorm} so that whenever the intention is normative in nature, the agent can generate in itself the emotional beliefs associated with compliance or non-compliance with the norm. After that, the agent will inform the rest of the society of the result of its actions (function \textit{.sendMsg}) and will execute all the necessary internal actions before restarting the reasoning cycle. 

\vspace{0.5cm}

\textbf{AffModB Step}
After having executed the ClrInt step, the agent will proceed to synchronize the beliefs between the normative rational cycle and the affective reasoning cycle using the fAffModB function.

\subsubsection{Affective Reasoning Cycle} \label{sec:affective_reasoning_cycle}
The affective reasoning cycle comprises the steps \textit{Appr, UpAs, SelCs, and Cope} in Figure \ref{fig:transition_states}. The agent begins its execution in an affective state of equilibrium, i.e., it does not feel any emotion with sufficient intensity to be considered meaningful.

The affective cycle starts in the step \textit{Appr} with the function \textit{fAppr}, where the appraisal process is performed on the basis of \textit{the concerns, personality and probabilities} represented as \textit{beliefs}. In addition, the function \textit{fAppr} determines whether an event is affectively relevant and generates the corresponding \textit{appraisal variables}. It then proceeds to the \textit{UpAs} step, where the \textit{fUpAs} function updates the affective state based on the \textit{appraisal variables} generated in the previous step. After updating the affective state, it proceeds to the step \textit{SelCs}, where the function \textit{fSelCs} checks if it is necessary to create new behaviors in the agent, and also verifies which \textit{coping strategies} are applicable. In the \textit{Cope} step, the \textit{fCope} function performs the necessary tasks to execute the corresponding \textit{coping strategies}. The intentions generated through these \textit{coping strategies} are added to the end of the agent's intention base, which contains the intentions of both the rational cycle and the affective cycle.

\subsubsection{\textit{Temporal Dynamics} Cycle}
This cycle, composed only by the step \textit{AsNrDecay}, contains a single function called \textit{fAsNrDecay}, which has been extended from the original proposal so that, in addition to promoting that the affective state tends to a state of equilibrium, the relevance of the rules is also reduced to reflect time passing. Thus, in the case of those norms for which a maximum compliance cycle is not specified, when they are not reinforced over time they will gradually lose relevance until, finally, they can be ignored by the agent, when their relevance is below a certain threshold defined by the developer.

\subsection{Transition Rules}\label{sec:transitionRules}
In this section we present only the new transition rules that present changes with respect to the work of Alfonso et al. \cite{alfonso2017toward}, shown in Figure \ref{fig:transition_states}, in order to integrate the normative concepts into the agent's reasoning cycle. 

\subsubsection{Perceive Step}
\textbf{Perceive}. This is the first step in the agent's reasoning cycle, in which it observes its environment to determine changes in its perceptions (\textit{PSet}, indicated in the diagram as \textit{Percepts}) through the evaluation function EvalP(PSet, $ag_{bs}$, $ag_{NB}$). \textit{NewP} contains the agent's new percepts based on its $ag_{bs}$ beliefs and $ag_{NB}$ normative beliefs, stored in its fact base, while \textit{RemP} contains all the removed $ag_{bs}$ and $ag_{NB}$ percepts.

\[
    \frac{EvalP(PSet, ag_{bs}, ag_{NB}) = \langle NewP,RemP \rangle}
         {\langle ag,C,M,T,Mem,Ta,Perceive,ast\rangle
          \rightarrow
          \langle ag',C,M,T,Mem,Ta',ProcMsg,ast\rangle}
\]

\begin{center}
\begin{tabular}{@{}lll@{}}
where: & $Ta'_{Ub}$     & $= \langle NewP, RemP, Perceive \rangle$
\end{tabular}
\end{center}

This transition rule makes use of the function \textit{EvalP}. The objective of this function is to compute the changes in the agent's perceptions. Where \textit{bs} is the set of beliefs of the agent, \textit{nbs} is the set of normative beliefs of the agent and \textit{PSet = \{pc,pc',...\}} is the set of observational perceptions in the environment per agent (\textit{pc} represents an individual perception, as a literal value). The set of new percepts, \textit{NewP}, is calculated as the difference between \textit{PSet$\setminus$agperc(bs,nbs)}. Similarly, the set of perceptions to be eliminated, \textit{RemP}, is calculated as the difference between \textit{PSet$\setminus$agperc(bs,nbs)} and \textit{PSet$\setminus$agperc(bs,nbs)}. The formal definition of the function is as follows:

\medskip
$EvalP(PSet,bs,nbs) = \langle NewP,RemP\rangle \mid NewP = \{bs \cup nbs \in PSet \mid bs\cup nbs\notin agperc(bs,nbs)\}$ and $RemP = \{bs \cup nbs \in agperc(bs,nbs) \mid bs\cup nbs\notin PSet\}$
\medskip

Meanwhile, the function \textit{agperc(bs,nbs)} allows, given a set of beliefs \textit{bs} and a set of normative beliefs \textit{nbs}, to obtain the set of perceptions of the agent as:

\medskip
$agperc(bs,nbs)=\{ bs[annot] \mid bs[annot] \in bs \cup nbs$ and $source(percept) \in annot \}$
\medskip

\subsubsection{ProcMsg Step}
The following transition rules correspond to the message reception process. These rules make use of the functions $S_M(M_{In})$, which allows the agent to choose a message from the queue of incoming messages $M_{In}$, and \textit{SocAcc(id, ilf, at, $NB_{assoc}$)}, which allows the agent to determine whether a message is socially acceptable, where \textit{id} is the message identifier, \textit{ilf} is the \textit{illocutionary force} of the message (i. e. \textit{Tell, Untell}), \textit{at} is the propsitional content of the message and finally $NB_{assoc}$ represents the normative belief that might be associated with the response (see Vieira et al. \cite{vieira2007formal} for more information about these functions).

\textbf{Receiving a Tell Message} When the agent is informed of a new norm, it must be able to create the necessary plans to respond to it. Therefore, after selecting a message that is socially acceptable, the agent will proceed to generate the normative plans using the function \textit{GenNormPlans}. Once these new plans have been generated, it will continue with its usual reasoning cycle.

\begin{align*}
    & \frac{
            \begin{aligned}
                S_M(M_{In}) = \langle mid,id,Tell,Bs\rangle \quad (mid,i) \notin M_{SI} \textrm{(for any intention i)} \\
                SocAcc(id,Tell,Bs)
            \end{aligned}
            }
            {\langle ag,C,M,T,Mem,Ta,ProcMsg,ast\rangle
              \rightarrow
              \langle ag',C',M',T',Mem,Ta,SelEv,ast\rangle}
\end{align*}

\begin{center}
\begin{tabular}{@{}ll@{}}
where: & $M'_{In} = M_{In} \setminus \{\langle mid, id, Tell, Bs\rangle\}$ \\
& and for each $b\in Bs:$ \\
& $ag'_{bs} = ag_{bs} + b[id]$ \\
& $ag'_{ps} = GenNormPlans(b[id], ag_{ps})$ \\
& $C'_E = C_E \cup \{\langle +b[id],\top \rangle \}$
\end{tabular}
\end{center}

The \textit{GenNormPlans} function, used to generate the possible normative plans based on the new messages, first evaluates whether the beliefs contained in the message are normative in nature, that is, whether they follow the pattern defined in the semantics of the language (see section \ref{sec:language_extension}). For example, the agent receives the following belief: \texttt{norm("prohibition", "np\_yell:at\_classroom", 0, 50, "ALL", [0.1, 0.1]).} which represents the prohibition of shouting while the agent is inside a classroom, with indefinite duration, which affects all the agents of the society and which in case of compliance is estimated to generate satisfaction in the agent and in the society. The agent decides to incorporate the rule and to do so modifies its list of plans, adding two new plans to its list. An example of the agent's criteria for creating plans can be seen in section \ref{sec:DemoScenario}. If the new beliefs are normative, the function returns the agent's list of plans concatenating the new normative plans, otherwise it returns the original list of plans:
 
\medskip
 \[ 
 GenNormPlans(b[id], ag_{ps}) = \begin{cases} 
        \mbox{b[id]:normative\_plan $\cup ag_{ps}$} & \mbox{if b[id] is norm} \\ 
                                                    & \mbox{and b[id]:normative\_plan$\in ag_{ps}$} \\ 
        \mbox{$ag_{ps}$} & \mbox{otherwise}
\end{cases} 
\]
\medskip

\textbf{Receiving a Tell Message as Reply} This relationship has been modified so that upon receiving a response that has an associated normative belief, the agent updates the relevance of the related norm: 

\begin{align*}
    & \frac{
            \begin{aligned}
                S_M(M_{In}) = \langle mid,id,Tell,Bs\rangle \quad (mid,i) \notin M_{SI} \textrm{(for any intention i)} \\
                SocAcc(id,Tell,Bs) \quad T_{a_{Ub}}=\langle NewP,RemP,Perceive\rangle
            \end{aligned}
            }
            {\langle ag,C,M,T,Mem,Ta,ProcMsg,ast\rangle
             \rightarrow
             \langle ag',C',M',T,Mem,Ta',AffModB,ast\rangle}
\end{align*}

\begin{center}
\begin{tabular}{@{}ll@{}}
where: & $M'_{In}$ = $M_{In}\setminus{\langle mid,id,Tell,Bs\rangle}$ \\
       & $M'_{SI}=M_{SI}\setminus{(mid,i)}$ \\
       & $C'_I = C_I \cup {i}$ \\
and for each b $\in$ Bs: & $ag_{bs}' = ag_{bs}'+{b[source(id)]}$ \\
       & $Ta'_{Ub} = \langle Bs', RemP, ProcMsg\rangle$ \\
       & if b[id] is a response to a norm: $ag'_{NB_{rel}}[b] = increment\_relevance(ag_{NB_{rel}}[b])$  \\
\end{tabular}
\end{center}

The increment\_relevance function is used to update the numerical value of the rule relevance. In this job the increment is fixed, for each response the agent receives to a normative action it has performed, the norm relevance is incremented by a $\delta$ value which in the default configuration is 0.1 but can be modified by the developer. 

\subsubsection{SelAppl Step}
\textbf{Selection of an Applicable Plan} This transition rule has been modified to reflect the changes made to the selection function $S_{O}$ which has two new parameters: the default value contained in the agent's personality $ag_{P_{rl}}$ and the set of normatively relevant beliefs $ag_{NB_{rel}}$. It is necessary to add these parameters to allow reordering by priority and consideration of compliance or non-compliance with norms mentioned in section \ref{sec:normative_reasoning_cycle}.

\[
    \frac{S_{O}(T_{Ap}, Ta_\sigma, ag_{P_{rl}}, ag_{NB_{rel}}) = (p, \theta)}
         {\langle ag, C, M, T, Mem, Ta, SelAppl, ast \rangle \rightarrow \langle ag, C, M, T', Mem, Ta, AddIM, ast \rangle}
\]
\begin{center}
    where: $T'\rho = (p, \theta)$
\end{center}

\subsubsection{SelInt Step}
\textbf{Intention selection} From the two possible transition rules (intention set has values or intention set is empty) the transition rule where the intention set has values has been modified to reflect changes in the selection function $S_{I}$, which has as a new parameter the set of normatively relevant beliefs $ag_{NB_{rel}}$. It is necessary to add this parameter to allow intentions associated with norms to have higher priority, as discussed in section \ref{sec:normative_reasoning_cycle}.

\[
    \frac{C_{I}\neq \{\} \qquad S_{I}(C_{I},ag_{NB_{rel}})=i}
         {\langle ag, C, M, T, SelInt \rangle \rightarrow \langle ag, C, M, T', ExecInt \rangle}
\]
\begin{center}
    where: $T_{\iota}' = i$
\end{center}

\subsubsection{ExecInt Step}
\textbf{Executing an intended means} When an action \textit{a}, which is contained in the body of a normative plan \textit{p}, is executed, the agent will execute the function ComplyToNorm(a, $ag_{NB}$) to check the compliance or non-compliance with the norm. This will lead to the generation of new affective beliefs that will be stored in \textit{Ta} to be processed in the AffModB step.

\[
    \frac{\begin{aligned}
              T_{i} = i[head \leftarrow a;h]
          \end{aligned} 
         }
         {\langle ag, C, M, T, Mem, Ta, ExecInt, ast \rangle \rightarrow \langle ag, C', M, T', Mem, Ta', AffModB, ast \rangle}
\]

\begin{center}
\begin{tabular}{@{}ll@{}}
where: & $C'_A = C_A \cup \{a\}$ \\
       & $T'_i = i[head\leftarrow h]$ \\
       & $C'_I = (C_I \setminus T_i) \cup \{T'i\}$ \\
and if a is norm: & $Ta'_{Ub} = \langle Bs'\cup ComplyToNorm(a, ag_{NB}), RemP, ProcMsg\rangle$ \\
\end{tabular}
\end{center}

The function ComplyToNorm(a, $ag_{NB}$) allows that, when the agent executes an action that is an obligation within the maximum term of compliance, the agent generates in itself a positive affective reaction and, when it executes an action that is a prohibition, it generates a negative affective reaction. These emotions are contained in the pre-appraisal of the normative belief:

\medskip
\[
ComplyToNorm(a, ag_{NB}) = \begin{cases}
        ag_{NB_{pa}} & a \in ag_{NB_{np}} and \\
                     & (ag_{NB_{do}} = \texttt{'obligation'}) and \\
                     & currentReasoningCycle < ag_{NB_{l}} \\
        opp\_emotion(ag_{NB_{pa}}) & a \in ag_{NB_{np}} and \\
                                   & (ag_{NB_{do}} = 'prohibition') and \\
                                   & currentReasoningCycle < ag_{NB_{l}} \\
        \varnothing  & otherwise
\end{cases}
\]
\medskip

$ag_{NB_{pa}}$ contains the emotion associated with compliance with the norm. As already discussed, the criterion for establishing which affective response generates compliance with the norm in the agent is positive. Therefore, if the agent executes an action associated with a prohibition (i.e. ignores the prohibition) a negative affective response must be generated. The opp\_emotion function calculates the inverse value of the pleasure and arousal values stored in the pre-appraisal. That is, if the pre-appraisal contained as values [-0.25,0.5] the vector returned by the function would be [0.25,-0.5].

\subsubsection{AsNrDecay Step}
\textbf{Affective-Normative temporal dynamic} In each cycle of AsNrDecay the agent recalculates the temporal decay of the affective state and of the relevance of the norms that have not been reinforced. For the calculation of the decay of the affective state, the function $AsDec(Ta_\sigma, ag_{P_{tr}})$ is used, where $Ta_\sigma$ represents the affective state and $ag_{P_{tr}}$ represents the agent's personality traits. For the calculation of norm relevance decay, the function $NrelDec(ag_{NB_{rel}}, Mem)$ is used, where $ag_{NB_{rel}}$ contains the current relevance of each normative belief and \textit{Mem} contains the temporal information of relevant events for the agent.

\[
    \frac{}{\langle ag,C,M,T,Mem,Ta,AsNrDecay,ast\rangle
            \rightarrow
            \langle ag',C,M,T,Mem,Ta',AsNrDecay,ast\rangle}
\]

\begin{center}
\begin{tabular}{@{}lll@{}}
where: & $ag'_{NB_{rel}}$ & $= NrelDec(ag_{NB_{rel}}, Mem)$ \\
       & $Ta'_\sigma$     & $= AsDec(Ta_\sigma, ag_{P_{tr}})$
\end{tabular}
\end{center}

\subsection{Language Extension} \label{sec:language_extension}
This section briefly describes the extensions made to the Jason language specification, extending the EBNF used in \cite{bordini2007programming} and \cite{alfonso2017toward}. The main extension within the agent framework is the component called \textit{norms}. In this way the developer can add rules at development time, as well as at runtime, thanks to the creation of new beliefs. For this purpose, the keyword \textit{norm}, which is used to indicate that a belief is normative, and the keyword \textit{np}, which is used to indicate that a plan is normative, have been defined. The normative \textit{belief} is composed of: a \textit{deontic operator}, which is a literal value used to indicate whether the rule corresponds to an obligation or a prohibition; a normative plan (identified by the reserved word \textit{np}), which contains the plan or plans that are affected by the rule; a \textit{limit\_cycle} containing a numerical value defining the maximum cycle in which the rule will be active; a \textit{relevance} that contains a numerical value to indicate the importance of the norm for the agent, this value increases when the agent performs an action affected by a norm and receives interactions from other agents and decreases when it receives no interactions; a list of roles that could be affected at the affective level by the fulfillment or non-fulfillment of the norm; a pre-appraisal, which contains the emotion that will be generated in the agent by the fulfillment of the norm, represented by a vector with numerical values, corresponding to the pleasure and arousal dimensions of the PAD model.

Furthermore, as we have already mentioned, the concept of rebelliousness (\textit{reb\_level})) has been added to the agent's personality, which is represented by a numerical value and indicates the extent to which the agent is willing to follow the rules.

Table \ref{tab:EBNF-mod} shows an extract with the EBNF syntax of the NEA agent language in Jason, where the terms added to the syntax proposed in \cite{alfonso2017toward}, with which the concept of norm is modeled, are highlighted in bold. Table \ref{tab:EBNF-exp} schematically describes each of these new terms, together with the values that each of them can take.

\begin{table}[ht]
\centering
\fontfamily{pcr}\selectfont
\begin{tabular}{lcl}

\underline{agent} & $\rightarrow$ &  \underline{init\_bels} \underline{init\_goals} \underline{plans} \\
& $\mid$ & \textbf{[} \underline{concerns} \textbf{]} \textbf{[} \underline{personality}
 \textbf{]}
 \textbf{[ \underline{roles} ]}
 \textbf{[ \underline{norms} ]} \\

\underline{personality} & $\rightarrow$ & "personality\_\_:" "\{" \underline{traits} \textbf{[} "," \underline{rat\_level} \textbf{]} \\ 
& $\mid$ & \textbf{[} "," \underline{coping\_strats} \textbf{]} \textbf{[} "," \underline{\textbf{reb\_level}} \textbf{]} "\}" "." \\

\underline{reb\_level} & $\rightarrow$ &  \textbf{<NUMBER>} \\

\underline{roles} & $\rightarrow$ & .\textbf{"roles\_\_:" "\{" [ \underline{role} ("," \underline{role})* ] "\}" "."} \\

\underline{role} & $\rightarrow$ & \textbf{<STRING>} \\

\underline{norms} & $\rightarrow$ & \textbf{"norms\_\_:" "\{" [ \underline{norm} ("," \underline{norm})* ] "\}" "."} \\

\underline{norm} & $\rightarrow$ & 
\textbf{"norm(" \underline{deontic\_operator} "," \underline{normative\_plan} ","} \\
& $\mid$ & \textbf{\underline{limit\_cycle} "," \underline{relevance} ","} \\
& $\mid$ & \textbf{\underline{list\_of\_affected\_roles} "," \underline{pre-appraisal} ")."} \\ 

\underline{deontic\_operator} & $\rightarrow$ &  \textbf{( "obligation" $\mid$ "prohibition" )} \\

\underline{normative\_plan} &  $\rightarrow$ & 
\begin{tabular}[t]{@{}l@{}}
    \textbf{[} "@" \underline{atomic\_formula} \textbf{]} \textbf{"np\_\_"} \underline{triggering\_event} \\
    \textbf{[} ":" \underline{context} \textbf{]} \\ 
    \textbf{[} "\texttt{<-}" \underline{body} \textbf{]}
\end{tabular} \\

\underline{limit\_cycle} & $\rightarrow$ &  \textbf{<NUMBER>} \\

\underline{relevance} & $\rightarrow$ &  \textbf{<NUMBER>} \\

\underline{list\_of\_affected\_roles} & $\rightarrow$ &  \textbf{( "ALL" $\mid$ ("," \underline{role})* )} \\

\underline{pre-appraisal} &$\rightarrow$& \textbf{[NUMBER, NUMBER]} \\

\end{tabular}

\caption{Simplified EBNF extension for NEA developing in Jason}
\label{tab:EBNF-mod}
\end{table}

\begin{table}[!ht]
\centering
\begin{tabular}{|l|l|p{\textwidth}|}
\hline
    & \textbf{Concept}  & \textbf{Component explanation}    \\ \hline
\multirow{7}{*}{\begin{tabular}[c]{@{}l@{}}Agent\\ EBNF\end{tabular}} & \multirow{2}{*}{personality} 
    & \textbf{reb\_level}: numerical value between [0.0,1.0] that indicates the extent to which the agent is willing to break the rules. The higher the value, the greater the probability that the agent will break the established rules. This component is optional.    \\ \cline{2-3} 
    & \multirow{5}{*}{norms}    & \textbf{norms}: The set of norms known to the agent. Each norm is described by its deontic operator, its associated plan (normative plan), its maximum cycle, its relevance, the set of roles that will be emotionally affected by this norm, and the emotion that will foreseeably be generated in the agent when fulfilling the norm. 
    \\ \cline{3-3} 
    &   & \textbf{deontic\_operator}: literal that allows defining whether the rule is a \textit{prohibition} or an \textit{obligation}. \\ \cline{3-3} 
    &   & \textbf{normative\_plan}: The normative plan that represents the agent's behavior in the event of compliance with the norm. \\ \cline{3-3} 
    &   & \textbf{limit\_cycle}: Numeric value defining the maximum cycle up to which the standard will be active, in the range [0, $\infty$[, where 0 indicates that the standard does not have a compliance limit cycle, i.e. it will always be active.\\ \cline{3-3} 
    &   & \textbf{relevance}: Numerical value that defines the relevance of the standard, in the range [0.0,$\infty$[. The higher the value, the higher the relevance. When \textit{relevance} falls below a certain value \textit{relevance\_threshold}, the rule will not be taken into account by the agent.. \\ 
    \cline{3-3} &   & \textbf{list\_of\_affected\_roles}: Set of roles that will be affected affectively by the fulfillment or non-fulfillment of the rule. It can take the value ALL if it affects the whole society, it can be left empty if the rule does not affect any role at the affective level or, finally, it can be a list of ROLES (ROLES being the name of each possible role). \\ \cline{3-3}
     &   & \textbf{pre-appraisal}: Estimation of the emotion that will be generated in the agent in the case of complying with the norm. The emotion is represented using the pleasure and arousal dimensions of the PAD model. Thus, they are represented using a vector [X,Y] where X and Y have values in the range [-1.0,1.0]. \\ \hline
    
\end{tabular}
\caption{Explanation of the new EBNF components for the NEA agent language in Jason}
\label{tab:EBNF-exp}
\end{table}

\section{Demo Scenario}
\label{sec:DemoScenario}
In order to demonstrate the feasibility of this proposal, the following example scenario has been designed in which a society of NEAs will have to interact and make use of the proposed mechanisms to regulate their behavior. After the COVID-19 pandemic, many universities are having to regulate and define a series of regulations that allow the gradual withdrawal of masks, since they are no longer indispensable and their use is once again optional. At the Polytechnic University of Valencia, in order to avoid sick leave and ensure that teaching can continue normally, the Rectorate has made it compulsory for the teaching staff to continue using face masks in the classroom. This measure has caused a disagreement between teachers who advocate the elimination of the mask and those who continue to defend its use, so that the rules conflict with the emotions caused by this measure in the individual. Outside the classroom, it is the teacher's choice to wear the mask. Students, on the other hand, are not required to wear masks in the classroom, but they are aware of the regulations imposed on teachers, so they expect them to wear them while they are in the classroom. Outside the classroom, on the other hand, they think that they are not necessary because they are outdoors and, therefore, although there is no rule determining this, they prefer teachers not to wear masks.

\begin{figure}[ht]
     \centering
     \begin{subfigure}[b]{0.45\textwidth}
         \centering
         \includegraphics[width=\textwidth]{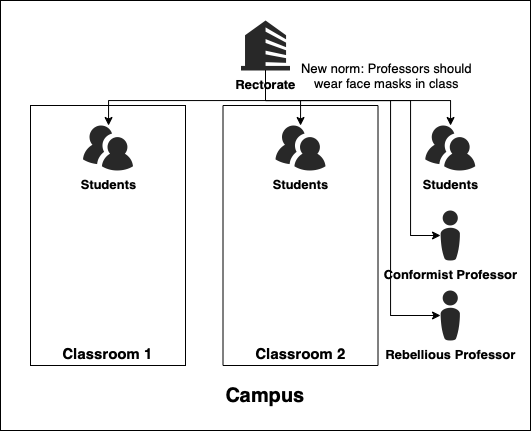}
         \caption{Rectorate send the new norm "Proffesors should wear face masks in classrooms" to the whole society}
         \label{fig:scenario1}
     \end{subfigure}
     \hfill
     \begin{subfigure}[b]{0.45\textwidth}
         \centering
         \includegraphics[width=\textwidth]{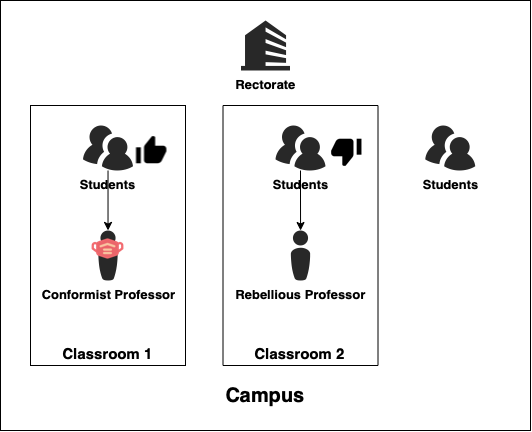}
         \caption{While respecting the institutional norm, the professor gets positive feedback, whereas the other gets negative}
         \label{fig:scenario2}
     \end{subfigure}
     \hfill
     \begin{subfigure}[b]{0.45\textwidth}
         \centering
         \includegraphics[width=\textwidth]{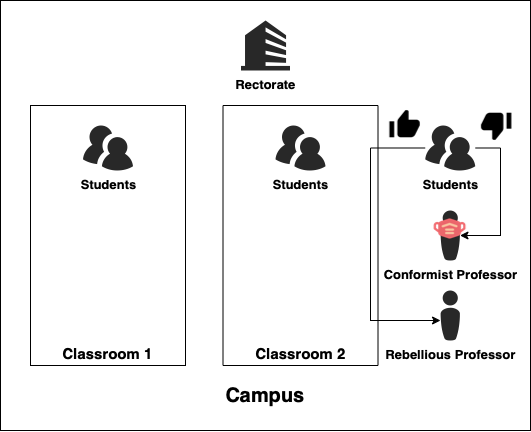}
         \caption{Students outside of classroom react positively to the professor respecting the social norm (not wearing mask), and negatively to the one who is not}
         \label{fig:scenario3}
     \end{subfigure}
    \hfill
    \begin{subfigure}[b]{0.45\textwidth}
         \centering
         \includegraphics[width=\textwidth]{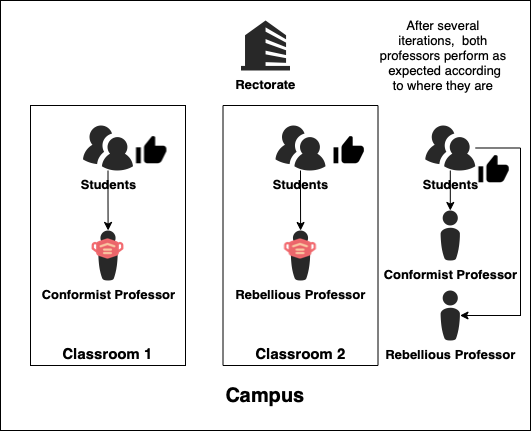}
         \caption{After several iterations receiving emotional feedback from the students, both professors respect both the institutional norm and the social norm}
         \label{fig:scenario4}
     \end{subfigure}
     \caption{Scenario schema}
     \label{fig:scenario-schema}
\end{figure}

In order to model this example, a society with different types of agents has been set up. Firstly, the agent \textit{Rectorate}, which is responsible for dictating and communicating institutional norms to the rest of the agents in the society. The agents with the role of \textit{Student}, who through observation of the professors give them emotional feedback on the impact on them of seeing that the professors respect the norms in the different areas of the university. Finally, the agents with the role of \textit{Professor}, which will be discussed in more detail in this example, will be in charge of going through the different areas of the university, where the students will be, each one behaving according to their personality, which will imply that they can decide not to use the mask in mandatory areas, and that based on their actions and the affective responses they receive, they will modify their behavior. Figure \ref{fig:scenario-schema} shows schematically the result of the execution of the experiment.

Here is the proposed configuration for the two agents with the role \textit{Professor} that will participate in the example. The first one represents a teacher with an introverted character, with a tendency to abide by the rules, who prefers to wear the mask at all times. The second represents an extroverted and rebellious teacher, who prefers to avoid wearing masks in order to feel closer to the students. In both cases, the Ôtextit{relevance} value of the agents is 25.0, i.e., all normative plans that have a relevance higher than 25.0 will be taken into account by the agent and will cease to have relevance in case the Ôtextit{relevance} value of the plan falls below that threshold. Initially, teachers have among their beliefs one that tells them where they are in the campus (i.e., campus, classroom) and they have two plans that allow them to switch between these two spaces (i.e., classroom, classroom). The initial content of the plan associated with the classroom entry is identical for the two types of teachers:\\
\texttt{+enter\_classroom:not in\_classroom <- -in\_campus; +in\_classroom; \\+teach\_lesson; -exit\_classroom.} \\
The initial content of the plan upon leaving class for the conformist teacher is:
\texttt{+exit\_classroom:in\_classroom <- -in\_classroom; +in\_campus; \\+enjoy\_freetime; +enter\_classroom.}\\
The initial contents of the plan upon leaving class for the wayward teacher are:
\texttt{+exit\_classroom:in\_classroom <- -in\_classroom; -wearing\_mask; \\ +in\_campus; +enjoy\_freetime; +enter\_classroom.}

At the beginning of the execution, the professors are outside the classrooms when the Rectorate communicates the new rule to the society with the message \texttt{norm("obligation", "np\_\_enter\_classroom:role(professor) \& \\ not wearing\_mask <- +wearing\_mask.", 0, 50.0, "ALL", [0.5,0.5])}.\\
This standard indicates that teachers who enter a classroom and are not wearing a mask are required to put on the mask. In addition, according to the content of the rule, it has an indefinite duration, the relevance value is above the agents' threshold, so it is of interest to them and, finally, it is indicated that the rule will have an affective impact on the whole society and that, in case of compliance, the affective state of the society will improve and worsen in the case of non-compliance.

When the professors process the rule, they proceed to create two new plans based on the original plan \texttt{enter\_classroom}, one for the case in which they decide to follow the rule, and one for the case in which they decide to skip it. In the case of the plan in which the agents decide to follow the norm, the header of the plan is identical to the one contained in the norm and the body of the plan concatenates the body of the original plan (in which it is indicated what the teacher does when entering the classroom) and the body of the plan contained in the norm to which, finally, the beliefs necessary to modify their affective state, contained in the pre-appraisal of the norm, are added. For the plan in which the agent decides to ignore the norm, the body of the new plan is added to the body of the original plan and the beliefs that modify its affective state as the inverse values of those contained in the pre-appraisal of the norm, ignoring in this case the body of the plan contained in the norm. Additionally, in order for the agents to choose which of the two plans to follow, to the head of both plans we add the result of some utility functions that are responsible for calculating the agent's intention. The plan with the higher score is the one chosen. For this example, the utility functions are calculated based on: 

\begin{itemize}
    \item The value of rebelliousness contained in the agent's personality. The higher the value, the less desire to follow the norm.
    \item The relevance value of the norm. The higher the value, the greater the desire to follow the norm.
    \item The current value of their affective state. The higher the value the less desire to follow the norm.
    \item The pre-appraisal value of the rule. If non-compliance modifies the agent's affective state in a negative way to a large extent (e.g. $new\_pleasure < old\_pleasure - 0.25$ or $new\_arousal < old\_arousal - 0.25$) the greater the desire to follow the norm.
    \item The percentage of agents to which the norm might be relevant, calculated on the basis of the number of agents who have the role(s) contained in the norm, ALL being 100 percent. The higher the percentage, the greater the desire to follow the norm.
\end{itemize}

Here are the functions employed to fulfill this tasks:

\texttt{comply = (1 - reb\_level) * \%ag\_affected * (aff\_state-new\_aff\_state) + relevance }

\texttt{break = reb\_level * (100 - \%ag\_affected) * (aff\_state+new\_aff\_state) - relevance}

Once the agents decide whether or not to follow the norm, they proceed to enter the classroom and execute the corresponding body of the plan. In case of following the norm they put on the mask and their affective state improves and in case of ignoring the norm they do not put on the mask and their affective state worsens. Subsequently, the agents with role \textit{Student} give feedback to the agents with role \textit{Professor} in relation to the compliance with the rule. 
This feedback is sent through messages containing an annotation with the reference to the norm that elicited the response and how compliance or non-compliance with the norm has affected them at the affective level. In this example, students improve their affective state when they see the norm fulfilled, and worsen it otherwise. 
When teachers receive the response and process it, they modify the relevance of the norm indicated in the message by increasing its value as \texttt{new\_norm\_relevance = max(old\_relevance + (0.1/number\_agents), 1)} and also modify their affective state. 
In case of positive affective responses as:

\texttt{new\_affective\_state = max(old\_affective\_state + (affective\_response / number\_agents), [1.0,1.0])}

and in case of negative affective responses as:

\texttt{new\_affective\_state = max(old\_affective\_state + (affective\_response / number\_agents), [-1.0,-1.0])}.

Initially, the conformist professor receives positive responses for following the norm, while the rebellious teacher receives negative responses for not following the norm. Eventually, when relevance's value increases enough, the rebelliousness value will be outweighed by the relevance value, so the rebellious professor will end up following the norm regardless of its rebelliousness value.

When professors go out on campus, there is no rule that determines whether or not they must wear a mask. Despite the fact students, who prefer that professor do not wear masks outdoors, will give affective responses to professor based on whether or not they wear masks. These responses contain a reference to the beliefs that elicited the affective response and the appraisal value. In the case of the rebellious teacher, by indicating his plan to remove his mask, he will receive positive affective responses, so he will only update his affective state following the formula already seen. In the case of the conformist teacher, who will leave the classroom without removing his mask, he will receive negative affective responses. These responses will be messages similar to: \texttt{"(+wearing\_mask;+in\_campus),[-0.1,-0.2]"} while wearing a mask outside. These beliefs are stored in the agent as a single belief, with the accumulated value of every identical belief registered. After several iterations in which identical beliefs are registered and when the accumulated resulting value deviates by more than [0.5,0.5] or [-0.5,-0.5], it will cause the agents to examine all their plans in search of those that have any relation with the content contained in the belief. In the case of finding any (i.e., after the execution of the plan, the agent could be in the state \texttt{in\_campus} and \texttt{wearing\_mask} simultaneously), the agent will proceed to create a copy of this plan to avoid reaching this undesired state. In the case of the rebellious professor this state is unreachable, since the only plan that allows him to be \texttt{in\_campus} is \texttt{exit\_classroom}, and in that plan he removes his mask before leaving. The same is not true for the conformist professor's \texttt{exit\_classroom} plan, in which the mask is not removed and thus could allow the new undesired state to be reached. For this reason, and in order to avoid accumulating sanctions, the agent will modify the body of the plan to include the removal of the mask, becoming the same as the rebellious professor's plan. In this way, the agents are aware of the social norm and modify their behaviour to ensure compliance with it.

\section{Conclusions and future work} \label{sec:Conclusions}
In this work we offer o formalisation for a Normative Emotional Agent (NEA), a General-purpose Intelligent Normative-Affective Agent Architecture, based on the $GenIA^3$ BDI agent architecture. This includes an extension of the AgentSpeak reasoning cycle, the definition of the modification required in its operational semantics, and the extension of the syntax of an AgentSpeak-based agent language (Jason) to include normative and affective attributes.

This allows the creation of intelligent agents that use both rules and emotions in their reasoning cycles. The main advantage of this is the possibility of replicating with greater fidelity the behavioral patterns of human societies in a computer system. To this end, when an agent identifies a norm, it modifies its agenda of plans to incorporate new plans that allow compliance or non-compliance with the norm. In addition, agents react affectively to compliance or non-compliance with norms, both for actions performed by themselves and for actions performed by other agents in society.

We are currently working on the reasoning process that allows the emergence of social norms. In this proposal we have simplified the process by using a system of sanctions and rewards that forces agents to eventually comply with the norms, based on the modifications that are made on their affective state.

\section{Acknowledgements}
This work was supported by the Spanish Government project PID2020-113416RB-I00, the TAILOR project, the Generalitat Valenciana project PROMETEO/2018/002 and the Spanish Goverment PhD Grant PRE2018-084940.

\bibliographystyle{abbrv}
\bibliography{bibliography}
\clearpage


\end{document}